\documentclass{aa}  

\usepackage{graphicx}
\usepackage{txfonts}

\newcommand{\feh}{\mbox{[Fe/H]}\xspace}
\newcommand{\teff}{\ensuremath{T_{\rm eff}}\xspace}
\newcommand{\logg}{\mbox{$\log g_*$}\,}
\newcommand{\vsini}{\mbox{$v \sin i_{*}$}\,}
\newcommand{\kms}{\mbox{km\,s$^{-1}$}}
\newcommand{\ms}{\mbox{m\,s$^{-1}$}}

\newcommand{\mjup}{\mbox{$\mathrm{M_{\rm Jup}}$}\xspace}
\newcommand{\rjup}{\mbox{$\mathrm{R_{\rm Jup}}$}\xspace}

\newcommand{\mstar}{\mbox{M$_{*}$}}
\newcommand{\rstar}{\mbox{R$_{*}$}}

\newcommand{\msol}{\mbox{$\mathrm{M_\odot}$}\xspace}
\newcommand{\rsol}{\mbox{$\mathrm{R_\odot}$}\xspace}

\newcommand{\lstar}{\mbox{$L_{*}$}\xspace}

\newcommand{\juliet}{{\sc \tt juliet}\xspace}
\newcommand{\radvel}{{\sc \tt radvel}\xspace}
\newcommand{\dynesty}{{\sc \tt dynesty}\xspace}
\newcommand{\exofast}{{\sc \tt EXOFASTv2}\xspace}

\usepackage{hyperref}
\begin{document}

    \title{An old warm Jupiter orbiting the metal-poor G-dwarf TOI-5542}
    \titlerunning{An old warm jupiter orbiting TOI-5542}
    \authorrunning{N. Grieves, et al.}
    
    \author{Nolan Grieves
    \inst{\ref{inst-geneva}}\fnmsep\thanks{\email{nolangrieves@gmail.com}}
    \and Fran\c{c}ois Bouchy \inst{\ref{inst-geneva}}
    \and Sol{\`e}ne Ulmer-Moll\inst{\ref{inst-geneva}}
    \and Samuel Gill\inst{\ref{inst-war},\ref{inst-warexo}}
    \and David~R.~Anderson\inst{\ref{inst-war},\ref{inst-warexo}}
    \and Angelica Psaridi \inst{\ref{inst-geneva}}
    \and Monika Lendl \inst{\ref{inst-geneva}}
    \and Keivan G. Stassun \inst{\ref{inst-vandy}}
    \and Jon M. Jenkins \inst{\ref{inst-ames}}
    \and Matthew R. Burleigh \inst{\ref{inst-leic}}
    \and Jack S. Acton \inst{\ref{inst-leic}}
    \and Patricia~T.~Boyd \inst{\ref{inst-goddard}}
    \and Sarah L. Casewell \inst{\ref{inst-leic}}
    \and Philipp Eigm\"uller\inst{\ref{dlr}}
    \and Michael R. Goad \inst{\ref{inst-leic}}
    \and Robert~F.~Goeke \inst{\ref{inst-mit}}
    \and Maximilian N. G\"unther \inst{\ref{estec}}\thanks{ESA Research Fellow}
    \and Faith Hawthorn \inst{\ref{inst-war},\ref{inst-warexo}}
    \and Beth A. Henderson \inst{\ref{inst-leic}}
    \and Christopher E. Henze \inst{\ref{inst-ames}}
    \and Andr\'es Jord\'an \inst{\ref{inst-uai},\ref{inst-mas}}
    \and Alicia Kendall \inst{\ref{inst-leic}}
    \and Lokesh Mishra \inst{\ref{inst-geneva}}
    \and Dan Moldovan \inst{\ref{inst-google}}
    \and Maximiliano Moyano \inst{\ref{inst-ucn}}
    \and Hugh Osborn \inst{\ref{inst-bern},\ref{inst-mit}}
    \and Alexandre Revol \inst{\ref{inst-geneva}}
    \and Ramotholo~R.~Sefako \inst{\ref{inst-saao}}
    \and Rosanna H. Tilbrook \inst{\ref{inst-leic}}
    \and St\'ephane Udry \inst{\ref{inst-geneva}}
    \and Nicolas Unger \inst{\ref{inst-geneva}}
    \and Jose I. Vines \inst{\ref{inst-uchile}}
    \and Richard~G.~West\inst{\ref{inst-war},\ref{inst-warexo}}
    \and Hannah L. Worters \inst{\ref{inst-saao}}
    }
    \institute{
    Observatoire de Gen{\`e}ve, Universit{\'e} de Gen{\`e}ve, 51 Chemin Pegasi, 1290 Versoix, Switzerland \label{inst-geneva}
    \and
    Department of Physics, University of Warwick, Gibbet Hill Road, Coventry, CV4 7AL, UK \label{inst-war}
    \and 
    Centre for Exoplanets and Habitability, University of Warwick, Gibbet Hill Road, Coventry, CV4 7AL, UK \label{inst-warexo}
    \and
    Vanderbilt University, Department of Physics \& Astronomy, 6301 Stevenson Center Lane, Nashville, TN 37235, USA \label{inst-vandy}
    \and 
    NASA Ames Research Center, Moffett Field, CA 94035, USA \label{inst-ames}
    \and
    School of Physics and Astronomy, University of Leicester, Leicester LE1 7RH, UK \label{inst-leic}
    \and
    Astrophysics Science Division, NASA Goddard Space Flight Center, Greenbelt, MD 20771, USA \label{inst-goddard}
    \and
    Institute of Planetary Research, German Aerospace Center, Rutherfordstrasse 2, 12489 Berlin, Germany\label{dlr}
    \and
    Department of Physics and Kavli Institute for Astrophysics and Space Research, Massachusetts Institute of Technology, Cambridge, MA 02139, USA \label{inst-mit}
    \and
    European Space Agency (ESA), European Space Research and Technology Centre (ESTEC), Keplerlaan 1, 2201 AZ Noordwijk, The Netherlands \label{estec}
    \and 
    Facultad de Ingenier\'ia y Ciencias, Universidad Adolfo Ib\'a\~nez, Av. Diagonal las Torres 2640, Pe\~{n}alol\'{e}n, Santiago, Chile \label{inst-uai}
    \and 
    Millennium Institute for Astrophysics, Chile \label{inst-mas}
    \and
    Google, Cambridge, MA, USA \label{inst-google}
    \and
    Instituto de Astronom\'ia, Universidad Cat\'olica del Norte, Angamos 0610, 1270709, Antofagasta, Chile \label{inst-ucn}
    \and
    Physikalisches Institut, University of Bern, Sidlerstrasse 5, 3012 Bern, Switzerland \label{inst-bern}
    \and
    South African Astronomical Observatory, P.O Box 9, Observatory 7935, Cape Town, South Africa \label{inst-saao}
    \and 
    Departamento de Astronom\'ia, Universidad de Chile, Casilla 36-D, Santiago, Chile \label{inst-uchile}
    }

   \date{Received 20 May 2022; accepted 19 September 2022}

 
  \abstract{We report the discovery of a 1.32$^{+0.10}_{-0.10}$ \mjup  planet orbiting on a 75.12 day period around the G3V $10.8^{+2.1}_{-3.6}$ Gyr old star TOI-5542 (TIC 466206508; TYC 9086-1210-1). The planet was first detected by the Transiting Exoplanet Survey Satellite (TESS) as a single transit event in TESS Sector 13. A second transit was observed 376 days later in TESS Sector 27. The planetary nature of the object has been confirmed by ground-based spectroscopic and radial velocity observations from the CORALIE and HARPS spectrographs. A third transit event was detected by the ground-based facilities NGTS, EulerCam, and SAAO. We find the planet has a radius of 1.009$^{+0.036}_{-0.035}$ \rjup and an insolation of 9.6$^{+0.9}_{-0.8}$ $S_{\oplus}$, along with a circular orbit that most likely formed via disk migration or in situ formation, rather than high-eccentricity migration mechanisms. Our analysis of the HARPS spectra yields a host star metallicity of \feh\,=\,$-$0.21\,$\pm$\,0.08, which does not follow the traditional trend of high host star metallicity for giant planets and does not bolster studies suggesting a difference among low- and high-mass giant planet host star metallicities. Additionally, when analyzing a sample of 216 well-characterized giant planets, we find that both high masses (4\,\mjup$<M_{p}<$\,13\,\mjup) and low masses (0.5\,\mjup$<M_{p}<$\,4\,\mjup), as well as both both warm (P\,$>$\,10 days) and hot (P\,$<$\,10 days) giant planets are preferentially located around metal-rich stars (mean \feh\,$>$\,0.1). TOI-5542b is one of the oldest known warm Jupiters and it is cool enough to be unaffected by inflation due to stellar incident flux, making it a valuable contribution in the context of planetary composition and formation studies.}

   \keywords{planets and satellites: detection, dynamical evolution and stability, formation, fundamental parameters, gaseous planets}

   \maketitle

%

\section{Introduction}

The first exoplanets discovered were Jupiter-sized planets with close-in orbits (period $<$ 10 days) around their host stars, known as hot Jupiters \citep[e.g.,][]{Mayor1995}. Even with a low occurrence rate of $<$1\% \citep[e.g.,][]{Zhou2019}, hot Jupiters remain one of the largest samples of known exoplanets due to the observational biases of current detection methods favoring close-in, large, and massive planets. Hot Jupiters suffer from intense stellar irradiation that can deposit energy into their interiors and cause these planets to have have radii larger than what would otherwise be expected based on internal structure models \citep[e.g.,][]{Guillot2002}. Hot Jupiters are also affected by powerful tidal forces that can lead to tidal locking and dampening of orbital eccentricity and the planetary rotation period (see \citealt{DawsonJohnson2018} for a review), as well as intense day-night contrasts \citep[e.g.,][]{Knutson2007}. Therefore, the original properties of hot Jupiters have been significantly altered by their environment since their formation, which hinders placing constraints on planet formation and evolution models from current observations.

Warm exoplanets, which we define as exoplanets with 10-200 day orbital periods, provide the opportunity to better understand planet formation and evolution as their atmospheres are less altered by their host star and their orbital arrangement reflects a less extreme migrational history, as compared to close-in planets. Previous studies have found giant planets ($M_{p}$ $>$ 100 M$_{\oplus}$ or 0.31 \mjup) on 10-100 day periods to be relatively less frequent than giant planets with shorter or longer orbital periods, known as the period valley e.g., \citep{Udry2003,Wittenmyer2010}. Transiting warm planets are particularly valuable because both the transit and radial velocity detection methods can be used to obtain accurate masses and radii of planets that have not been exposed to strong atmospheric escape, which is essential for constraining the initial planet atmospheric mass fraction \citep[e.g.,][]{Kubyshkina2019a,Kubyshkina2019b}.

The  space-based all-sky Transiting Exoplanet Survey Satellite \citep[TESS;][]{Ricker2015} currently in operation has discovered thousands of transiting exoplanet signals. TESS first surveyed the entire sky in 26 different sectors, each 24 degrees by 96 degrees across, and viewed each sector for $\sim$27 days. Although some stars, such as in the ecliptic poles, are in overlapping sectors that allow for longer baseline coverage, most stars only have $\sim$27-day segments of continuous coverage. Planets with periods longer than $\sim$27 days can therefore only exhibit a single transit in one sector and their periods cannot be determined without more measurements. However, when combined with ground-based follow-up observations, TESS single transits have led to the discovery of new warm giant planets \citep[e.g.,][]{Gill2020c,Ulmer-Moll2022} as well as low-mass eclipsing binaries \citep[e.g.,][]{Lendl2020,Gill2020a,Gill2020b,Gill2022}. Low-mass eclipsing binaries are crucial for testing stellar evolution models of low-mass stars \citep{Kraus2011,Feiden2015} that are need to obtain accurate calculations of the masses and radii of single low-mass stars and of any exoplanets they may host.

Here, we report the discovery and characterization of a warm Jupiter first detected by TESS as a single transit event. We detail our observations in Section \ref{sec:obs}, including our radial velocity follow-up observations. In Section \ref{sec:analysis}, we describe our analysis to derive both stellar and planet properties and we present our results. We discuss our results in Section \ref{sec:discussion} and we summarize our conclusions in Section \ref{sec:conclusion}.

\section{Observations} \label{sec:obs}

\begin{figure}
  \centering
  \includegraphics[width=0.49\textwidth]{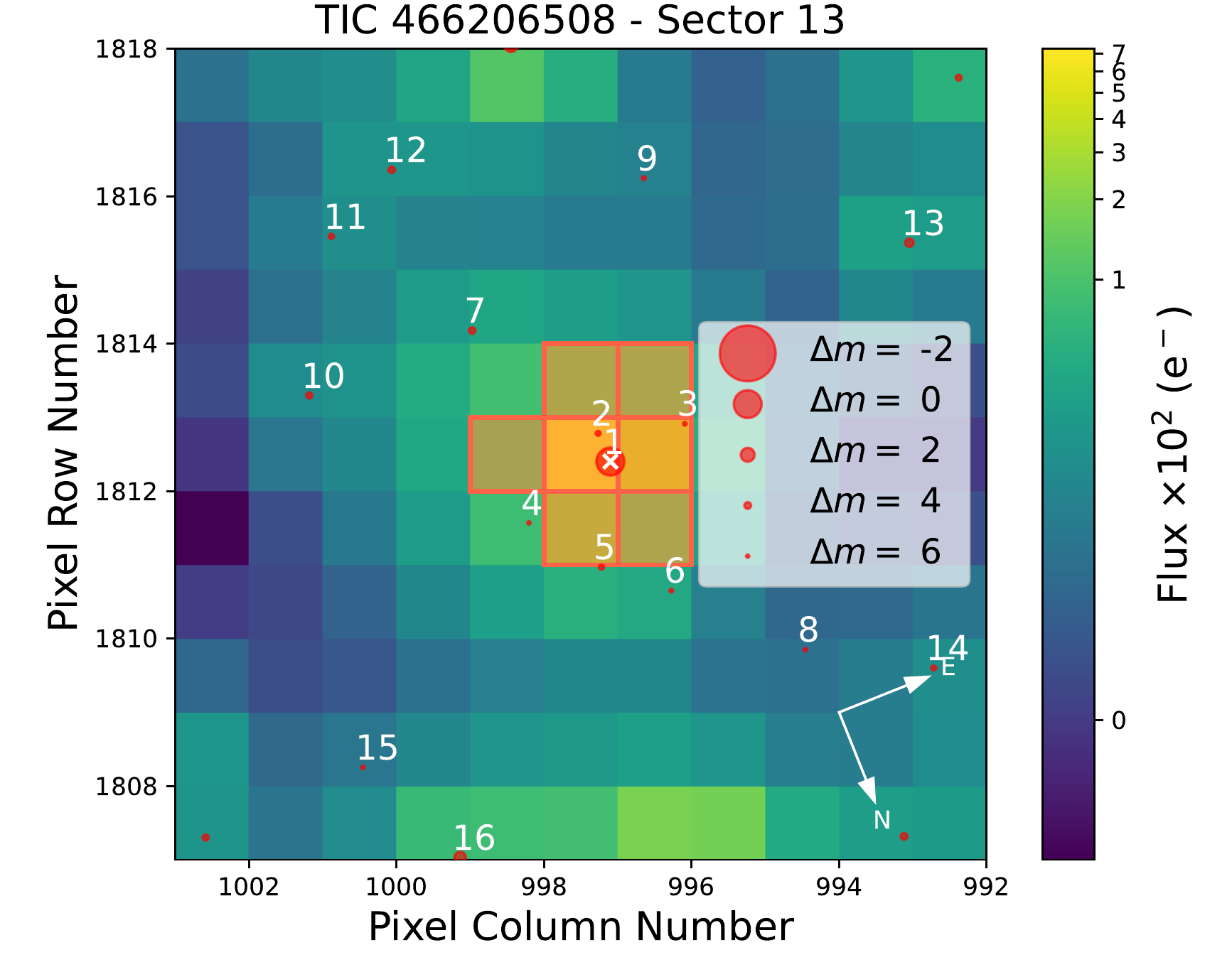}
    \includegraphics[width=0.49\textwidth]{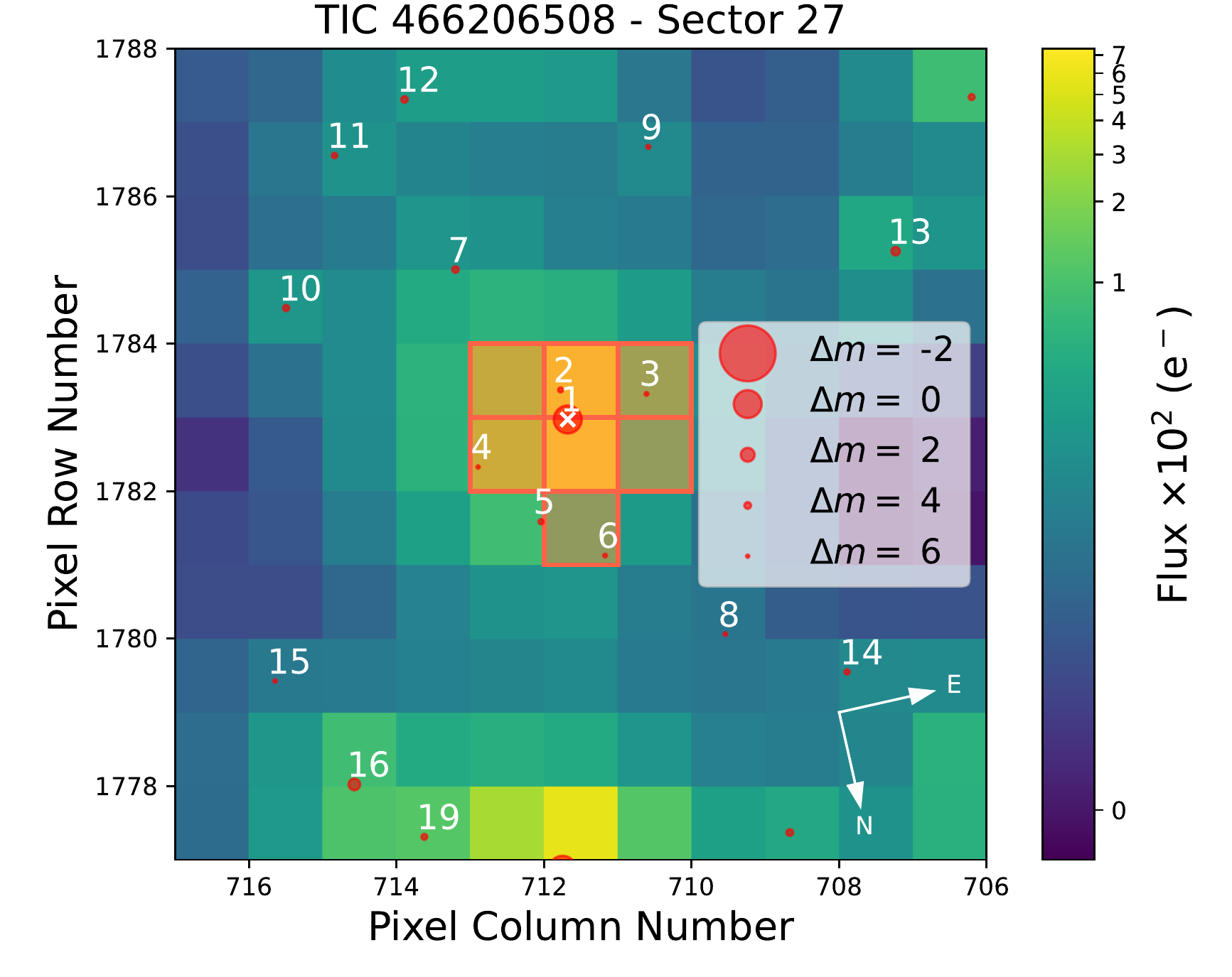}
 \caption{Gaia EDR3 catalog over-plotted on the TESS Target Pixel File of TOI-5542 (TIC 466206508) for Sector 13 (top) and Sector 27 (bottom).}
  \label{fig:tpfplot}
\end{figure}

\subsection{Observations and follow-up summary}

Our team searched through TESS light curves for single-transit events as described in \citet{Gill2020a,Gill2020b,Gill2020c}. For our search and preliminary analysis of the TESS transits, we used a custom transit model with a least squares fit utilizing a power-2 limb-darkening law presented by \citet{Maxted2019} and described in \citet{Gill2020a}. For our initial fit, we fixed the orbital period to 30 days and fit only the transit epoch, $T_0$, the scaled orbital separation, $R_1/a$, the ratio of radii, $k = R_2/R_1$, the impact parameter, $b$, and the photometric zero-point, $zp$. We identified single transits in the TESS data of TOI-5542 in Sectors 13 and 27 that were $T_{2-1}$\,$\approx$\,375.628 days apart. After inspecting the full light curves from TESS sectors 13 and 27, we identified the period of the planet to be greater than that of the 27-day sectors. This left possible aliases of $\approx$\,187.8, 125.2, 93.9, 75.1, 62.6, 53.7, 47.0, 41.7, 37.6, 34.1, 31.3, and 28.9 days. We subsequently began spectroscopic and radial velocity (RV) follow-up of the target with the CORALIE spectrograph. We also started photometric follow-up on NGTS to detect any possible transit events from the ground. 

\begin{figure}
  \centering
  \includegraphics[width=0.47\textwidth]{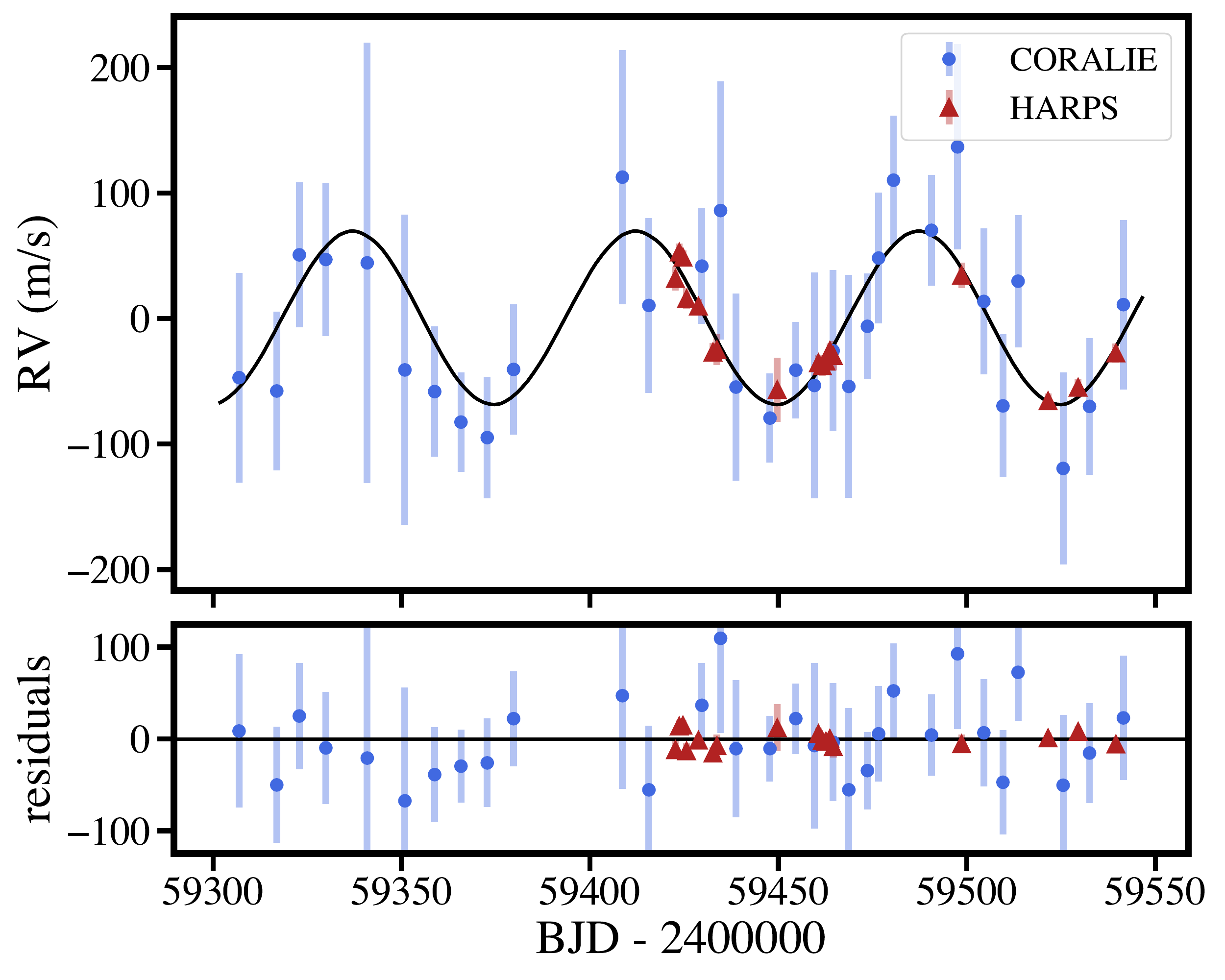}
  \includegraphics[width=0.47\textwidth]{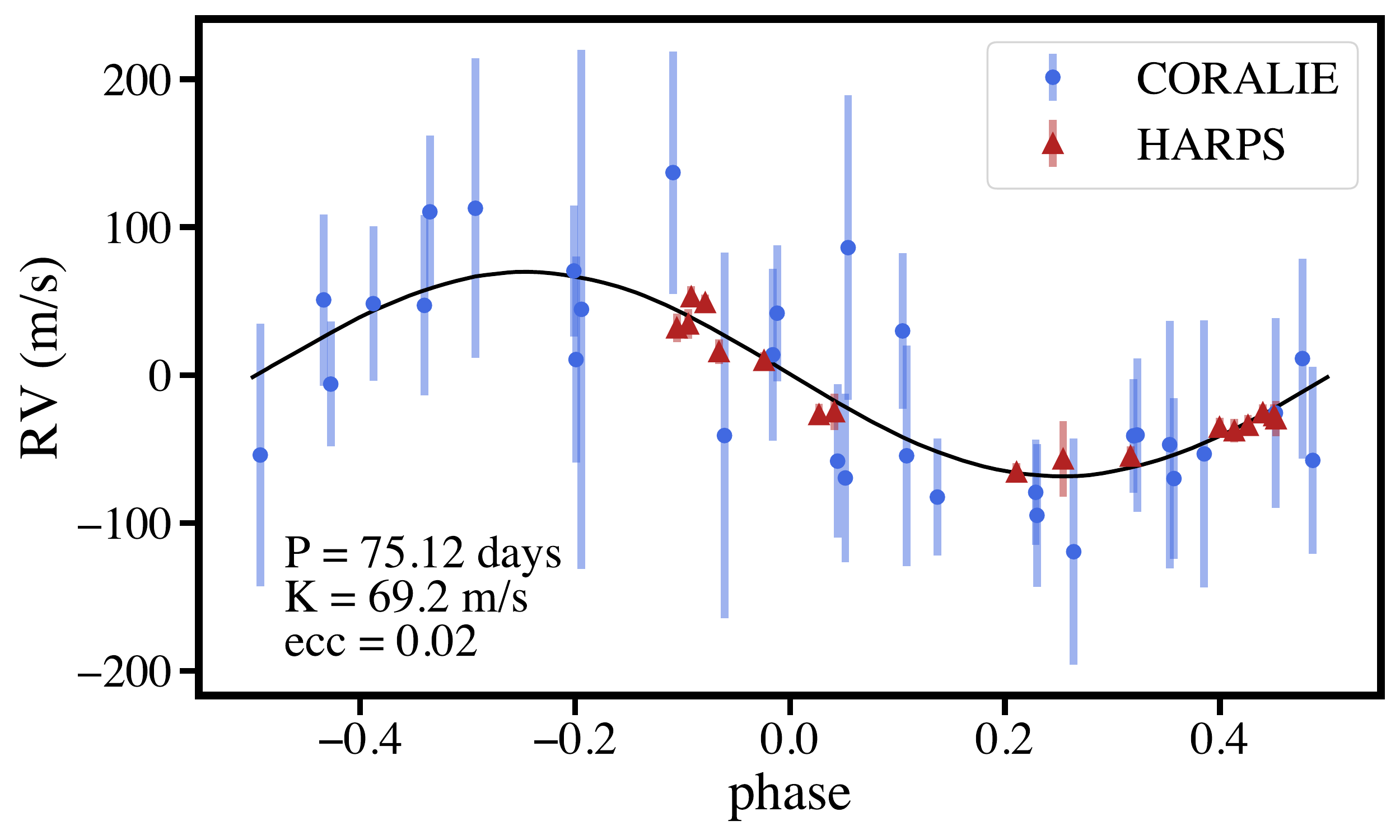}
  \includegraphics[width=0.47\textwidth]{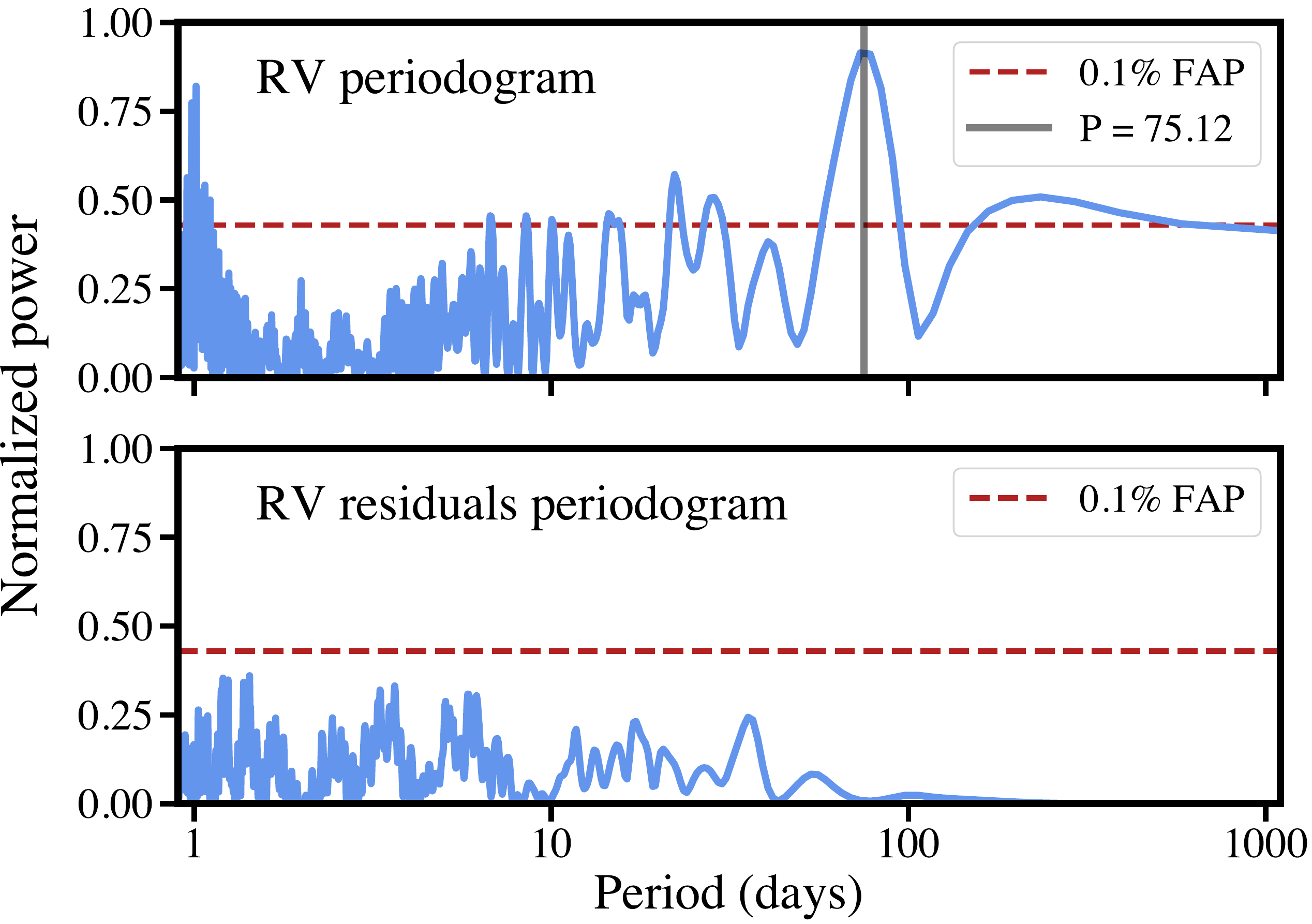}
  \caption{CORALIE and HARPS RVs, as described in Section \ref{sec:obs}. Top plot shows the full time baseline of the RV observations, while the middle plot shows the RVs phased to the 75.12 day period. The black lines show the best-fit Keplerian model to the RVs from our \juliet analysis, as described in Section \ref{sec:planetanalysis}. Lower plot shows the Lomb-Scargle periodogram of the original data as well as the residuals. Red dashed lines show a false alarm probability (FAP) of 0.1\%.}
  \label{fig:rvs}
\end{figure}

After 12 CORALIE observations, the highest peak in a periodogram analysis of the RVs was at 68 days -- closest to the 75.1 and 62.6 day aliases. We continued observing with CORALIE and began observing the target with the HARPS spectrograph. The HARPS RVs confirmed the 75.1 day alias as the most likely period, which is evident in the periodogram in Figure \ref{fig:rvs}. Finally, we identified on the night of August 3, 2021 that all alias periods would exhibit a transit, as this date was two times the length of time between the two TESS transits ($T_{2-1}\times$2\,$\approx$\,751.3 days after the TESS Sector 13 transit). We therefore scheduled the target on SAAO, EulerCam, and NGTS to capture the transit on August 3, 2021. We obtained partial transits on EulerCam and NGTS due to the target rising in the Western hemisphere and a partial transit on SAAO due to the night ending in South Africa. We did not catch an ingress or egress with NGTS due to scheduling constraints. We detail the observations for each facility below.

\subsection{Space-based TESS photometry}

TOI-5542 was observed by TESS in Sector 13 (22 June 2019 to 17 July 2019 UTC) with full-frame image (FFI) observations at a cadence of 30 minutes. We identified a single-transit event in TOI-5542 in our search of the Sector 13 data that was clearly significant compared to the out-of-transit data; with our preliminary transit fit, we calculated a signal-to-noise (S/N) of\,$\sim$\,35 by dividing the transit depth by the median absolute deviation of the light curve. For our analysis of the Sector 13 data, we used the FFI 30-minute light curve that was processed by the Science Processing Operations Center \citep[SPOC;][]{Jenkins2016} pipeline, which we obtained from the Barbara A. Mikulski Archive for Space Telescopes (MAST) astronomical data archive hosted by the Space Telescope Science Institute. The SPOC pipeline was first applied to TESS 2 minute data and then later applied to the 30-minute FFI data \citep{Caldwell2020}. The SPOC pipeline produces two light curves per sector, referred to as  simple aperture photometry (SAP) and presearch data conditioning simple aperture photometry \citep[PDCSAP;][]{Smith2012,Stumpe2012,Stumpe2014}. We used the PDCSAP light curves for our analysis. 

\begin{figure}
  \centering
  \includegraphics[width=0.49\textwidth]{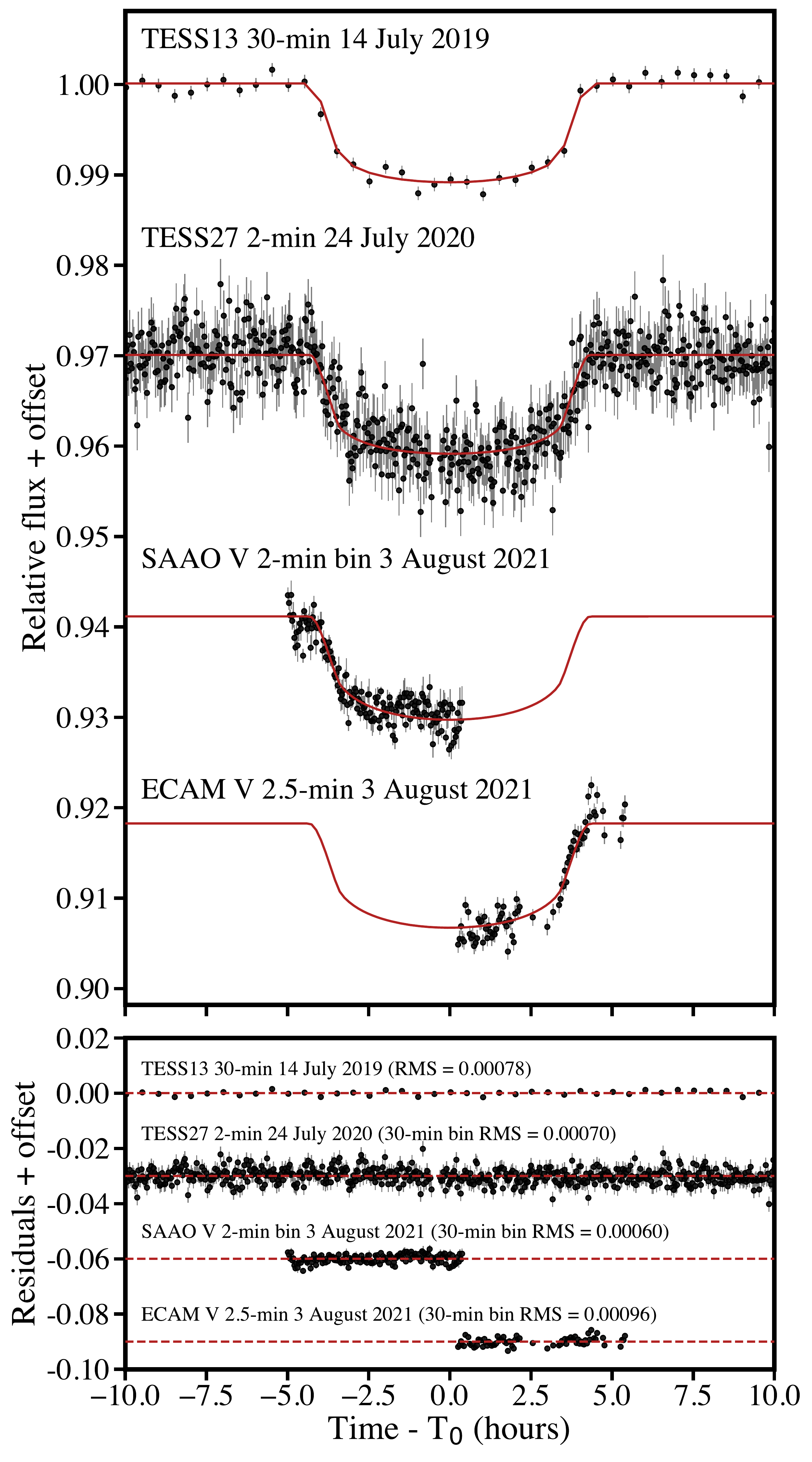}
 \caption{\textit{}Light curves around each transit of TOI-5542b from the respective facilities, as described Section \ref{sec:obs} (top). The SAAO data is binned to 2 minutes for visual purposes. The red lines show the best-fit transit model to each photometry data set from our \juliet analysis, as described in Section \ref{sec:planetanalysis}. TESS Sectors 13 and 27 data have the Gaussian process component of their models removed. Residuals of each transit fit and the root mean square (rms) of the residuals for each data set (bottom). The rms values for TESS27, SAAO, and EulerCam are calculated with the data binned to 30 minutes for comparison.}
  \label{fig:transits}
\end{figure}

We plotted the TESS target pixel file of Sector 13 in Figure \ref{fig:tpfplot} using \texttt{tpfplotter}\footnote{\url{https://github.com/jlillo/tpfplotter}} \citep{Aller2020} to show the possible contamination from known sources in the $Gaia$ Early Data Release 3 (EDR3) catalog \citep{GaiaCollaboration2021}, which we discuss further in Section \ref{sec:planetanalysis}. To account for stellar activity as well as residual instrumental systematics, we applied a Gaussian process \citep[e.g.,][]{Gibson2014,Haywood2014} to the TESS Sector 13 data (discussed further in Section \ref{sec:planetanalysis}). In Figure \ref{fig:transits}, we display the TESS Sector 13 data during the transit event with the Gaussian process model removed . The full TESS Sector 13 light curve with the Gaussian process model is displayed in Figure \ref{fig:TESS13_full}.

TOI-5542 was also observed by TESS in Sector 27 (5 July 2020 to 29 July 2020 UTC) with 2-minute cadence exposures\footnote{TOI-5542 obtained 2-minute data in Sector 27 as a result of the approved TESS Guest Investigator program G03188 PI: Villanueva}. We also searched TESS Sector 27 light curves for single-transit events and identified another event around TOI-5542 with a S/N\,$\sim$\,33, which we identified to be similar in shape and depth as the single-transit event found in Sector 13. For our analysis, we used TESS 2-minute data that were processed by the SPOC pipeline and again we used the PDCSAP light curves for our analysis. As in Sector 13, we also accounted for stellar activity and residual instrumental systematics using a Gaussian process for the TESS Sector 27 data (detailed further in Section \ref{sec:planetanalysis}). We display the TESS Sector 27 data during the transit event with the Gaussian process model removed in Figure \ref{fig:transits}. The full TESS Sector 27 light curve with the Gaussian process model is displayed in Figure \ref{fig:TESS27_full}.

\subsection{NGTS photometry}

After identifying the transit events in the TESS data we began monitoring TOI-5542 with the  Next  Generation  Transit  Survey  \citep[NGTS;][]{Wheatley2018}, a ground-based photometric survey located in Chile at the European Southern Observatory (ESO) Paranal site. NGTS consists of twelve fully robotic telescopes each with a 20\,cm photometric aperture and a wide 8 deg$^{2}$ field-of-view. It uses a custom filter spanning 520 to 890 nm and routinely achieves photometric precision of 150 parts per million. It has discovered a 3.18 R$_{\oplus}$ transiting planet \citep[NGTS-4b;][]{West2019}, which was the shallowest transiting system ever discovered from the ground. \citet{Wheatley2018} gives the full details of the NGTS facility and project.

TOI-5542 was observed using between one and five cameras per night for a total of 137 nights during the time baseline of UT 14 April 2021 to 7 April 2022 with an exposure time of 10 seconds and cadence of 13 seconds. The reduction of raw NGTS data is detailed in \citet{Bryant2020} and uses a custom aperture photometry pipeline that utilizes SEP \citep{Bertin1996,Barbary2016}. The pipeline computes photometry for a set of circular apertures with a range of radii and identifies the aperture size with the minimum root mean square (rms) scatter to create the final light curve. For TOI-5542, we used an aperture radius of four pixels. Comparison stars are ranked to select the best non-saturated non-variable comparison stars according to their color, brightness, and position on the image, relative to the target star. To avoid night-to-night and camera offsets, the same comparison stars were used for each camera and each night, and we normalized the data for each camera separately by dividing by its median flux value.

For our analysis, we removed large outliers in flux and flux uncertainty and binned the data to 30 minute bins. We display the full NGTS light curve as well as a Lomb-Scargle periodogram of the data in Figure \ref{fig:ngts_full}. We did not find a significant signal in the periodogram around the expected rotation period from the HARPS spectral analysis ($P_{\rm rot}$ / $\sin i_{*}$ = 17.7\,$\pm$\,3.0; see Section \ref{sec:star}). We obtained NGTS data on the night beginning  August 3, 2021, when a transit on TOI-5542 occurred, but we did not observe the ingress or egress of the transit. The offset between this night and those around it has an amplitude expected from the planet's transit. However, given the offsets of the entire data set the dip in the flux on this night could originate from telescope systematics, a variable companion star (which, however, was not detected), or from the host star itself. Therefore, we do not include the NGTS data in our global analysis of the planet.

\subsection{CORALIE spectroscopy and radial velocities}

After detecting two single transit events in the TESS data of TOI-5542, we initiated spectroscopic follow-up observations and radial velocities (RVs) of the star with the CORALIE spectrograph on the Swiss 1.2 m Euler telescope at La Silla Observatory, Chile \citep{Queloz2001}. CORALIE has a resolution of $R$\,$\sim$\,60,000 and is fed by two fibers: a 2 arcsec on-sky science fiber encompassing the star and another fiber that can either connect to a Fabry-P\'erot etalon for simultaneous drift correction or on-sky for background subtraction of sky flux. We observed TOI-5542 in the simultaneous Fabry-P\'erot drift correction mode. We obtained 31 CORALIE observations between 2021 April 2 to 2021 November 23 with 2400 second exposures. We reduced the spectra with the standard calibration reduction pipeline and computed RVs by cross-correlating with a binary G2 mask \citep{Pepe2002}. We obtained typical RV uncertainties of $\sim$58 \ms\, for our CORAIE RVs. The CORALIE RVs are displayed in Figure \ref{fig:rvs} and presented in Table \ref{tab:rv}

\subsection{HARPS spectroscopy and radial velocities}

Following the initial RV observations with CORALIE, we began observing TOI-5542 with the HARPS spectrograph \citep{Pepe2002,Mayor2003} to obtain a more precise mass measurement and more precise orbital properties. HARPS is hosted by the ESO 3.6-m telescope at La Silla Observatory, Chile and has a resolving power of $R$\,$\sim$\,115,000. TOI-5542 was observed with two HARPS programs\footnote{TOI-5542 was observed with HARPS programs Bouchy: 105.20L0 and Ulmer-Moll: 108.22L8} that both have the goal of characterizing long-period transiting giant planets. We obtained 18 HARPS observations of TOI-5542 from 2021 July 27 to 2021 November 21 with exposure times of 1800 seconds. The RVs were computed using the standard data reduction pipeline with a binary G2 mask. We obtained typical RV uncertainties of $\sim$7.5 \ms\,for our HARPS RVs. We did not include the HARPS RV observation on the night of  August 3, 2021 in our planetary analysis (Section \ref{sec:planetanalysis}) as TOI-5542b was transiting on that night and the RV would likely be affected by the Rossiter-McLaughlin effect as a result \citep{Rossiter1924,McLaughlin1924}. We predicted the Rossiter-McLaughlin effect measured with the classical method is $\sim$26\,\ms\,for TOI-5542 (Eq. 40 from \citealt{Winn2010}). 

The HARPS spectra were also used to derive spectral parameters for TOI-5542, as detailed in Section \ref{sec:star}. The HARPS RVs are displayed in Figure \ref{fig:rvs} and presented in Table \ref{tab:rv}. Figure \ref{fig:rvs} also displays the best-fit Keplerian model (see Section \ref{sec:planetanalysis}) for both the CORALIE and HARPS RVs, the RVs phased at the best-fit period of the planet, and the Lomb-Scargle periodogram \citep{Lomb1976,Scargle1982} of the combined CORALIE and HARPS RVs as well as the Lomb-Scargle periodogram of the RV residuals after subtracting the best-fit Keplerian model from the RVs. The original periodogram clearly shows the strong $\sim$75 day signal of the planet, while the residuals do not have any significant signals.

\subsection{EulerCam photometry}

We also observed TOI-5542 on the night of August 3, 2021 with EulerCam on the 1.2 m Euler telescope at La Silla Observatory, Chile. The observations were carried out for 5.14 hours allowing for the observation of an egress of the transit. We used a V filter and 150 second exposures for the observations. For details on the instrument and the data analysis routines used to extract relative aperture photometry with EulerCam, we refer to \citet{Lendl2012}. The EulerCam (ECAM) light curve is displayed in Figure \ref{fig:transits}. A full table of reduced EulerCam photometric observations is available in a machine-readable format at the CDS.

\subsection{SAAO photometry}

TOI-5542 was also observed the night of August 3, 2021 on the 1-m telescope at the South African Astronomical Observatory (SAAO) in Sutherland, South Africa. The observations were taken with one of the Sutherland High-speed Optical CCD Cameras (specifically "SHOCnAwe", Plate scale = 0.167 arcsec/pixel, binned to $4\times4$ pixels) in the V filter with 20-second exposures for 5.39 hours allowing us to catch a full ingress of the transit. Conditions were clear with somewhat variable seeing from $<1.5$ to $2.5$\,arcseconds. The data were reduced and calibrated with the local SAAO SHOC pipeline \citep{Coppejans2013}, which is driven by {\sc python} scripts running {\sc iraf} tasks ({\sc pyfits} and {\sc pyraf}). Aperture and differential photometry was performed using the {\sc Starlink} package {\sc autophotom}, utilizing a single comparison star and an aperture radius of four pixels was selected to maximize the signal-to-noise. This telescope and instrument setup has been used in previous studies to detect transits of planets, brown dwarfs, and low-mass stars \citep[e.g.,][]{Smith2021,Acton2021,Gill2022}. The SAAO photometry data binned to 2 minutes for display purposes is shown in Figure \ref{fig:transits}. A full table of reduced SAAO photometric observations is available in a machine-readable format at the CDS.

\section{Analysis and results} \label{sec:analysis}

\subsection{Stellar parameters}\label{sec:star}

\begin{table}
\caption{\label{tab:stellar} Stellar parameters of TOI-5542.}
\resizebox{\columnwidth}{!}{%
        \begin{tabular}{lcc}
        \hline\hline
        \noalign{\smallskip}
        Parameter       &       Value   &       Source\\

        \hline
    \noalign{\smallskip}
    \noalign{\smallskip}
    \multicolumn{3}{l}{\underline{Identifying Information}}\\
    \noalign{\smallskip}
    \noalign{\smallskip}
    TYC ID & TYC 9086-01210-1 &  Tycho \\
    TESS ID & TIC 466206508 &  \textit{TESS} \\
    2MASS ID &  2MASS J20111163-6108076  &  2MASS \\
    Gaia ID &  6443054270942726144 & \textit{Gaia} EDR3 \\
    
    \\
    \multicolumn{3}{l}{\underline{Astrometric parameters}}\\
    \noalign{\smallskip}
    \noalign{\smallskip}
    R.A. (J2000, h:m:s)         &        20:11:11.62    &  \textit{Gaia} EDR3    \\
        Dec      (J2000, h:m:s)         &        -61:08:07.68   &  \textit{Gaia} EDR3    \\
    Parallax  (mas) & 2.827 $\pm$ 0.030 & \textit{Gaia} EDR3 \\
    Distance  (pc) & 353.7 $\pm$ 3.7 & \textit{Gaia} EDR3 \\
    \\
    \multicolumn{3}{l}{\underline{Photometric parameters}}\\
    \noalign{\smallskip}
    \noalign{\smallskip}
        B               & 13.063 $\pm$ 0.020    & APASS \\ 
        V               & 12.402 $\pm$ 0.032    & APASS \\
        g       & 12.863 $\pm$ 0.030 & APASS \\
        r       & 12.244 $\pm$ 0.030 & APASS \\
        i       & 12.087 $\pm$ 0.030 & APASS \\
    G           & 12.272 $\pm$ 0.020    & \textit{Gaia} EDR3 \\
    B$_P$               & 12.601 $\pm$ 0.020    & \textit{Gaia} EDR3 \\
    R$_P$               & 11.779 $\pm$ 0.020    & \textit{Gaia} EDR3 \\
    J           & 11.266 $\pm$ 0.022    & 2MASS \\
        H               & 10.945 $\pm$ 0.023    & 2MASS \\
        K$_S$       & 10.897 $\pm$ 0.021        & 2MASS \\
    W1      & 10.819 $\pm$ 0.030        & WISE \\
    W2      & 10.870 $\pm$ 0.030    & WISE \\
    W3      & 10.738 $\pm$ 0.088    & WISE \\
    A$_{V}$     & 0.046 $\pm$ 0.036 & Sec. \ref{sec:star} \\
    \\

  \multicolumn{3}{l}{\underline{Bulk parameters}}\\
    \noalign{\smallskip}
    \noalign{\smallskip}
    \teff\,(K) & $5700\pm80$   & Sect. \ref{sec:star}\\
    \feh (dex) & $-0.21\pm0.08$ & Sect. \ref{sec:star}\\
    \logg (cm\,s$^{-2}$) & $4.2\pm0.2$   & Sec. \ref{sec:star}\\
    Spectral type & G3V & Sect. \ref{sec:star}\\
        $v \sin i_{*}$ (km\,s$^{-1}$) & $3.03 \pm 0.50$         & Sect. \ref{sec:star}\\
        $P_{\rm rot}$ / $\sin i_{*}$ (days) & $17.7 \pm 3.0$    & Sect. \ref{sec:star} \\
    Mass ($M_{\odot}$) & $0.890^{+0.056}_{-0.031}$  & Sect. \ref{sec:star} \\
    Radius ($R_{\odot}$) & $1.058\pm0.036$  & Sect. \ref{sec:star} \\
    $\rho_*$ (g\,cm$^{-3}$) & $1.07\pm0.13$  & Sect. \ref{sec:star}\\
    Luminosity ($L_{\odot}$) & $1.057\pm0.093$  & Sect. \ref{sec:star} \\
        Age     (Gyrs) & $10.8^{+2.1}_{-3.6}$ & Sec. \ref{sec:star} \\
        log(R'$_{\rm{HK}}$) & $-$5.28 $\pm$ 0.26 & Sec. \ref{sec:activity} \\
        \noalign{\smallskip}
        \hline
        \noalign{\smallskip}
    \end{tabular}}

 \label{tab:star_table}
\end{table}   

Obtaining proper and precise constraints on host star parameters is an essential component of exoplanet studies, as many critical properties of the planet directly depend on the estimated stellar properties, such as planet mass, radius, and insolation. There are several methods used to constrain various stellar properties, including spectral analysis, approximating the spectral energy distribution (SED) by integrating fluxes from broadband photometry in combination with model atmospheres \citep[e.g.,][]{vanBelle2009,StassunTorres2016}, as well as stellar isochrones and evolutionary models \citep{Yi2001,Dotter2016}, which essentially constrain the stellar mass and radius based on the stellar effective temperature, \teff, metallicity, \feh, and surface gravity, \logg (see Section 2 of \citet{Eastman2019} for a detailed discussion). The \teff and \feh are typically best constrained by spectral analysis \citep[e.g.,][]{Stassun2017,Jofre2019}. The \logg can also be estimated by spectral analysis, but often not as accurately \citep[e.g.,][]{Torres2012}. The SED fit  can typically best constrain the stellar radius, \rstar,\ and stellar luminosity, \lstar, by combining the broadband photometry and parallax information \citep[e.g.,][]{StassunTorres2016}, whereas evolutionary models are likely best suited to constrain the stellar mass, \mstar,\,and age \citep{Tayar2022}.

\subsubsection{HARPS spectra analysis}

We first used the HARPS spectra of TOI-5542 to determine \teff, \logg, \feh, and the rotational broadening projected into the line of sight, \vsini. Spectra were co-added onto a common wavelength axis to increase the signal-to-noise ratio (S/N) prior to spectral analysis. We found the HARPS co-added spectrum to have a S/N of $\sim$54. We used \texttt{ISPEC} \citep{Blanco-Cuaresma2014} to synthesize models using the radiative transfer code \texttt{SPECTRUM} \citep{Gray1999}, \texttt{MARCS} model atmospheres \citep{Gustafsson2008}, and version 5 of the $GAIA$ ESO survey (GES) atomic line list within \texttt{ISPEC} with solar abundances from \citet{Asplund2009}. The temperature was determined by fitting the iron lines and H-$\alpha$ lines, which were well matched to find \teff=\,5700\,$\pm$\,80\,K. We used the Mg triplets and Na doublet lines to determine \logg=\,4.2\,$\pm$\,0.2 and we found a metallicity of \feh=\,$-$0.21\,$\pm$\,0.08 using the individual FeI and FeII lines. There is no evidence of lithium, which suggests that the star is on the main sequence branch. Finally, from the HARPS spectra, we calculated a stellar rotational velocity of \vsini\,=\,3.03\,$\pm$\,0.50 km\,s$^{-1}$, which indicates a moderately slow rotator. With this value, we can put an upper limit on the stellar rotation period of $P_{\rm rot}$ / $\sin i_{*}$ = 2$\pi$\rstar/\vsini = 17.7\,$\pm$\,3.0\,days using the \rstar\, value derived in Section \ref{sec:SED}. We use these as our final \teff, \logg, and \feh values presented in Table \ref{tab:star_table}. We find the star to be a G3V spectral type from its \teff value using the updated table from \citet{PecautMamajek2013} \footnote{\url{https://www.pas.rochester.edu/~emamajek/EEM_dwarf_UBVIJHK_colors_Teff.txt}}.

As an independent check, we also derived the stellar parameters with the co-added HARPS spectra using SpecMatch-Emp \citep{Yee2017}, which uses a large library of stars with well-determined parameters to match the input spectra and derive spectral parameters. We use a spectral region that includes the Mg I b triplet (5100 - 5400 $\AA$) to match our spectra. SpecMatch-Emp uses $\chi^{2}$ minimization and a weighted linear combination of the five best matching spectra in the SpecMatch-Emp library to determine stellar parameters. With SpecMatch-Emp, we found similar stellar parameters, as compared to our other method with \teff = 5701\,$\pm$\,110\,K, \logg = 4.43\,$\pm$\,0.12, \feh = $-$0.18\,$\pm$\,0.09, \mstar\,$\sim$\,0.95\,\msol, \rstar\,$\sim$\,1.01\,\rsol, and an age of $\sim$\,9.59\,Gyr. However, we used the parameters derived from the first spectral analysis method above as our final spectral atmospheric parameters since we consider these values to be more precise based on the fact that the estimates of the atmospheric parameters are obtained by modeling individual spectral lines, rather than integrated quantities of entire spectra. However, we note that both methods are in good agreement.

\subsubsection{Spectral energy distribution analysis} \label{sec:SED}

\begin{figure}
  \centering
  \includegraphics[width=0.45\textwidth]{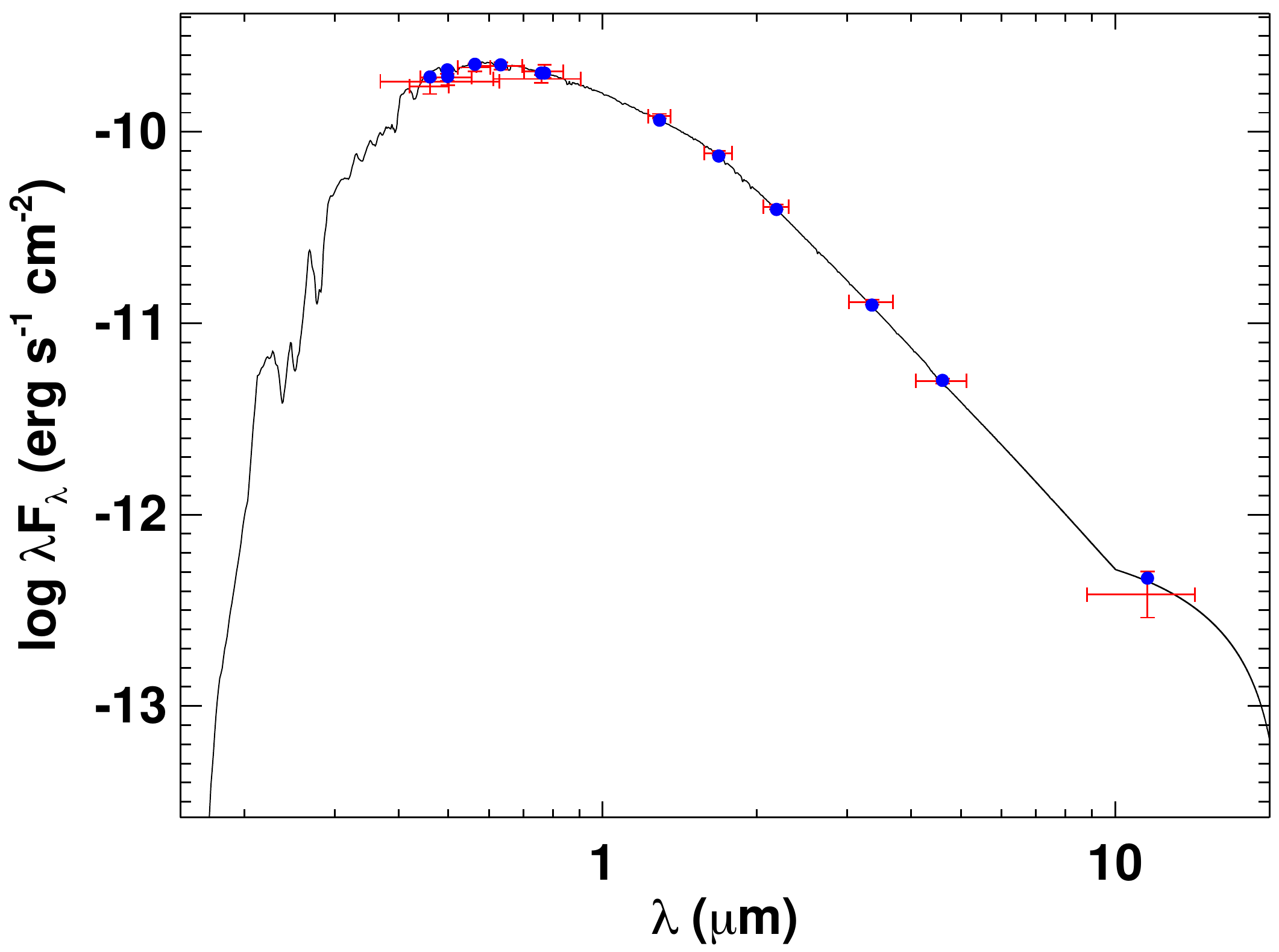}
  \caption{Spectral energy distribution (SED) for TOI-5542. Red symbols represent the observed photometric measurements, where the horizontal bars represent the effective width of the passband. Blue symbols are the model fluxes from the best-fit Kurucz atmosphere model (black).}
  \label{fig:sed_keivan}
\end{figure}

We performed a SED fit using Kurucz stellar atmosphere models \citep{Kurucz2013} with the $Gaia$ parallax following the procedures described in \citet{StassunTorres2016,Stassun2017,Stassun2018}, with the priors on \teff, \feh, and \logg from the spectroscopic analysis, and the extinction ($A_V$) limited to the maximum line-of-sight extinction from the Galactic dust maps of \citet{Schlegel1998}. We put a Gaussian prior for parallax from the value and uncertainty in $Gaia$ EDR3 \citep{GaiaCollaboration2021}. We note the $Gaia$ EDR3 parallax (2.8042) was  corrected by subtracting $-$0.0232 mas according to the \citet{Lindegren2021} prescription. For the SED fit we use photometry from APASS DR9 $BVgri$ \citep{Henden2016}; $Gaia$ EDR3 G, B$_P$, and R$_P$ \citep{GaiaCollaboration2021}; 2MASS J, H, and K$_S$ \citep{Skrutskie2006}; and ALL-WISE W1, W2, and W3 \citep{Wright2010}, which are presented in Table \ref{tab:star_table}. The SED fit gives our final stellar radius value of \rstar\,=\,1.058\,$\pm$\,0.036\,\rsol and is displayed in Figure \ref{fig:sed_keivan}. The SED also gives the V-band extinction, $A_{V}$, and the stellar luminosity presented in Table \ref{tab:star_table}.

\subsubsection{Stellar evolutionary model} \label{sec:star_mist}

We used the Modules for Experiments in Stellar Astrophysics (MESA) Isochrones \& Stellar Tracks \citep[MIST;][]{Choi2016,Dotter2016} isochrones within the \exofast \citep{Eastman2013,Eastman2017,Eastman2019} exoplanet and star modeling suite, which has the option of solely fitting the star. This suite explores the parameter space through a differential evolution Markov Chain coupled with a Metropolis-Hastings Monte Carlo sampler, and a built-in Gelman-Rubin statistic \citep{GelmanRubin1992,Gelman2003,Ford2006} is used to check the convergence of the chains. We use the \teff, \logg, and \feh values along with their uncertainties from the HARPS spectral analysis as priors. We did not simultaneously fit the SED so as not to bias the stellar radius from models and we relied on the more empirical SED-only fit above. We therefore used our SED fitted \rstar\, value as a prior in the MIST model. This MIST fit gives our final stellar mass and age, which are given in Table \ref{tab:star_table}.

\begin{table*}
\centering
\begin{minipage}{16cm}
\caption{TOI-5542 parameters from \juliet: median and 68\% confidence interval.}
\begin{tabular}{lllc}\hline 

\hline\hline

\noalign{\smallskip}
Parameter       & &    Prior distribution\raggedright* & Value     \\
\hline
\smallskip\\\multicolumn{2}{l}{\underline{Modeled instrumental parameters:}}&\smallskip\\

~~~~$q_{\text{1,TESS}}$\dotfill & Quadratic limb-darkening parametrization\dotfill & $\mathcal{N}(0.335,0.011)$\dotfill & $0.339^{+0.010}_{-0.010}$ \\
~~~~$q_{\text{2,TESS}}$\dotfill & Quadratic limb-darkening parametrization\dotfill & $\mathcal{N}(0.261,0.030)$\dotfill & $0.277^{+0.027}_{-0.028}$ \\
~~~~$m_{\text{flux,TESS13}}$\dotfill & Offset (relative flux)\dotfill & $\mathcal{N}(0,0.01)$\dotfill & $-0.00011^{+0.00018}_{-0.00025}$ \\
~~~~$\sigma_{\text{GP,TESS13}}$\dotfill & GP amplitude (relative flux)\dotfill & $\mathcal{J}(10^{-6},1)$\dotfill &  $0.00044^{+0.00033}_{-0.00012}$ \\
~~~~$\rho_{\text{GP,TESS13}}$\dotfill & GP time-scale (days)\dotfill & $\mathcal{J}(10^{-6},10^{3})$\dotfill &  $2.30^{+2.51}_{-1.00}$ \\
~~~~$m_{\text{flux,TESS27}}$\dotfill & Offset (relative flux)\dotfill & $\mathcal{N}(0,0.01)$\dotfill &  $-0.00009^{+0.00010}_{-0.00010}$ \\
~~~~$\sigma_{\text{GP,TESS27}}$\dotfill & GP amplitude (relative flux)\dotfill & $\mathcal{J}(10^{-6},1)$\dotfill &  $0.000432^{+0.000074}_{-0.000056}$ \\
~~~~$\rho_{\text{GP,TESS27}}$\dotfill & GP time-scale (days)\dotfill & $\mathcal{J}(10^{-6},10^{3})$\dotfill &  $0.54^{+0.20}_{-0.12}$ \\
~~~~$m_{\text{flux,SAAO}}$\dotfill & Offset (relative flux)\dotfill & $\mathcal{N}(0,0.01)$\dotfill &  $-0.00117^{+0.00016}_{-0.00017}$ \\
~~~~$\sigma_{\text{SAAO}}$\dotfill  & Jitter (ppm)\dotfill & $\mathcal{J}(0.1,10000)$\dotfill & $1585.4^{+144.7}_{-146.0}$ \\
~~~~$m_{\text{flux,ECAM}}$\dotfill & Offset (relative flux)\dotfill & $\mathcal{N}(0,0.01)$\dotfill &  $-0.00819^{+0.00021}_{-0.00021}$ \\
~~~~$\sigma_{\text{ECAM}}$\dotfill  & Jitter (ppm)\dotfill & $\mathcal{J}(0.1,10000)$\dotfill & $1096.7^{+162.6}_{-161.0}$ \\
~~~~$\mu _{\text{CORALIE}}$\dotfill & Systemic RV offset (\kms)\dotfill & $\mathcal{U}(-100,100)$\dotfill & $-54.4042^{+0.0101}_{-0.0105}$ \\
~~~~$\sigma_{\text{CORALIE}}$\dotfill  & Jitter (\ms)\dotfill &  $\mathcal{J}(0.01,200)$\dotfill & $0.50^{+4.66}_{-0.46}$ \\
~~~~$\mu_{\text{HARPS}}$\dotfill & Systemic RV offset (\kms)\dotfill &  $\mathcal{U}(-100,100)$\dotfill & $-54.3751^{+0.0025}_{-0.0026}$ \\
~~~~$\sigma_{\text{HARPS}}$\dotfill  & Jitter (\ms)\dotfill &  $\mathcal{J}(0.01,200)$\dotfill & $5.31^{+3.10}_{-3.08}$ \\

\smallskip\\\multicolumn{2}{l}{\underline{Modeled physical parameters:}}&\smallskip\\

~~~~$P$ \dotfill & Period (days) \dotfill & $\mathcal{U}$(75.12$\pm$0.2)\dotfill & $75.12375^{+0.00019}_{-0.00018}$ \\
~~~~$T_0$ \dotfill & Time of transit center (BJD$_{\text{TDB}}$) \dotfill & $\mathcal{U}$(2458679.3$\pm$0.2)\dotfill & $2458679.3476^{+0.0015}_{-0.0015}$ \\
~~~~$K$\dotfill & Radial velocity semi-amplitude (\ms) \dotfill & $\mathcal{U}$(0,1000)\dotfill & $69.17^{+4.34}_{-4.55}$ \\
~~~~$e$\dotfill & Eccentricity of the orbit \dotfill & $\mathcal{B}$(0.867,3.03)\dotfill & $0.018^{+0.026}_{-0.013}$ \\
~~~~$\omega$\dotfill & Argument of periastron (deg) \dotfill & $\mathcal{U}$(0,360)\dotfill & $125.6^{+171.4}_{-82.7}$ \\
~~~~$r_1$\dotfill & Parametrization for p and b \dotfill & $\mathcal{U}$(0,1)\dotfill & $0.6126^{+0.0388}_{-0.0503}$ \\
~~~~$r_2$\dotfill & Parametrization for p and b \dotfill & $\mathcal{U}$(0,1)\dotfill & $0.0980^{+0.0010}_{-0.0010}$ \\
~~~~$\rho _{*}$\dotfill & Stellar density (g\,cm$^{-3}$)  \dotfill & $\mathcal{N}$(1.07,0.13)\dotfill & $1.03^{+0.11}_{-0.10}$ \\

\smallskip\\\multicolumn{2}{l}{\underline{Derived planetary parameters$^{**}$:}}&\smallskip\\

~~~~$i$\dotfill & Inclination (deg) \dotfill & \dotfill & $89.643^{+0.069}_{-0.061}$ \\
~~~~$p=R_p/R_{\star}$\dotfill & Planet-to-star radius ratio \dotfill & \dotfill & $0.0980^{+0.0010}_{-0.0010}$ \\
~~~~$b$\dotfill &Impact parameter of the orbit \dotfill & \dotfill & $0.419^{+0.058}_{-0.075}$ \\
~~~~$a$\dotfill & Semi-major axis (AU)  \dotfill & \dotfill & $0.332^{+0.016}_{-0.016}$ \\
~~~~$M_p$\dotfill & Planetary mass (\mjup)  \dotfill & \dotfill & $1.32^{+0.10}_{-0.10}$ \\
~~~~$R_p$\dotfill & Planetary radius (\rjup)  \dotfill & \dotfill & $1.009^{+0.036}_{-0.035}$ \\
~~~~$\rho _p$\dotfill & Planetary density (g\,cm$^{-3}$)  \dotfill & \dotfill & $1.60^{+0.22}_{-0.19}$ \\
~~~~$S$\dotfill & Insolation ($S_{\oplus}$)  \dotfill & \dotfill & $9.6^{+1.0}_{-0.9}$ \\
~~~~$T_{eq}$\dotfill & Equilibrium Temperature ($K$) \dotfill & \dotfill & $441^{+48}_{-74}$ \\

\hline
\end{tabular}
\label{tab:planet}
{\raggedright*$\mathcal{U}(a,b)$ indicates a uniform distribution between $a$ and $b$; $\mathcal{J}(a,b)$ a Jeffrey or log-uniform distribution between $a$ and $b$; $\mathcal{N}(a,b)$ a normal distribution with mean $a$ and standard deviation $b$; and $\mathcal{B}(a,b)$ a Beta prior as detailed in \citet{Kipping2014}. \\ 
\raggedright**We sampled from a normal distribution for the stellar mass, stellar radius, and stellar temperature, based on the results from Section \ref{sec:star} to derive the parameters. \\
}
\end{minipage}
\end{table*}

\subsection{Joint RV and transit fit} \label{sec:planetanalysis}

We obtained the planetary parameters by jointly modeling the photometric and RV data with \juliet \citep{Espinoza2019}, which uses Bayesian inference to model a set number of planetary signals using {\sc \tt batman}\xspace \citep{Kreidberg2015} to model the planetary transit and \radvel \citep{Fulton2018} to model the RVs. For the light curves stellar activity as well as instrumental systematics can be taken into account with Gaussian processes \citep[e.g.,][]{Gibson2014} or simpler parametric functions. Several RV instruments can be taken into account with RV offsets fit between them. For the transit model, \juliet performs an efficient parameterization by fitting for the parameters $r_{1}$ and $r_{2}$ to ensure uniform exploration of the $p$ (planet-to-star ratio; $R_p/R_{\star}$) and $b$ (impact parameter) parameter space. For our joint RV and transit global model, we used the nested sampling method \dynesty \citep{Speagle2019} implemented in \juliet with 1000 live points and we ran the fit until the estimated uncertainty on the log-evidence was smaller than 0.1.

Table \ref{tab:planet} displays all of the modeled parameters as well as their input priors for our joint RV and transit global model. We used a uniform prior of 75.12$\pm$0.2 days for the period of the planet from the RV-identified alias of the two transits and a uniform prior of 0.2 days from the TESS Sector 13 transit for the time of transit center. We put broad uniform priors on the RV semi-amplitude up to 1000\,\ms\,and broad uniform priors of $\pm$100\,\kms\,for the systematic RV offsets of the CORALIE and HARPS instruments. From the near sinusoidal curve of the RV measurements, the planet is likely in a circular orbit; however, when allowing for the eccentricity as well as $e\cos{\omega}$ and $e\sin{\omega}$ to be freely constrained, \juliet was not converging,  as it was likely stuck in a nonphysical parameter space. Therefore, we added a Beta prior for transiting planets, as described in \citet{Kipping2014}. We used the $\rho_*$ (stellar density) as a parameter instead of the scaled semi-major axis ($\it{a}$/\rstar). The normal prior on stellar density is informed by the stellar analysis in Section \ref{sec:star}, which allowed us to derive a precise mass and radius along with their associated errors. 

For each photometric filter, we derived the quadratic stellar limb-darkening coefficients and their uncertainties using the {\sc \tt LDCU}$\footnote{\url{https://github.com/delinea/LDCU}}$ code, a modified version of the python routine implemented by \citet{EspinozaJordan2015}. The code uses two libraries of stellar atmosphere models, ATLAS9 \citep{Kurucz1979} and PHOENIX \citep{Husser2013}, to compute stellar intensity profiles for any given instrumental passband. The atmosphere models are selected based on a given set of stellar atmospheric parameters. The uncertainties on the stellar parameters are propagated by selecting several models within the uncertainty range and weighting them accordingly. The obtained intensity profiles are then fit with linear, square-root, quadratic, three-parameter, non-linear, logarithmic, exponential, and power-2 laws. The code performs three different fits for each intensity profile and each law and provides a series of additional coefficient values for each law by merging the outcomes from the previous fits. The merging is done based on the assumption of normal distributions from the estimated uncertainties, with the merging process taking place before we then recompute the global uncertainties from quantiles. This approach allows us to compute a precise median value while also allowing the uncertainties to encompass the whole merged distribution and remain more conservative. For the TESS data, we used the calculated limb-darkening coefficients of $q_1$\,=\,0.335$\pm$0.011 and $q_2$\,=\,0.261$\pm$0.030 as Gaussian priors in our global analysis. We fixed the calculated limb darkening parameters for the SAAO and EulerCam ground-based photometry due to the lower precision of the data. We determined values of $q_1$\,=\,0.521$\pm$0.010 and $q_2$\,=\,0.325$\pm$0.024 for the V bandpass used for the SAAO and EulerCam photometry.

We added a white-noise jitter term in quadrature to the error bars of both the photometry and RV data to account for underestimated uncertainties and additional noise that was not captured by the model. The jitter terms were fit using large log-uniform priors. For both TESS sectors 13 and 27 photometry, we found the jitter terms to be $<$0.1 parts-per-million (ppm) and not varying significantly from 0, so we fixed the jitter parameter to 0 for both TESS light curves. As displayed in Figure \ref{fig:tpfplot}, the TESS light curves have a few faint stars (G $>$ 17) in the aperture. However, the PDCSAP light curves are corrected for contamination from nearby stars by analyzing a synthetic scene constructed from the pointing model, the pixel response functions (PRFs), and the TIC catalog \citep{Bryson2010,Bryson2020}. For sector 27 the crowding metric (CROWDSAP) was 0.99635, indicating that the catalog predicts that 99.64\% of the flux in the photometric aperture is from TOI-5542 and for sector 13 the CROWDSAP is 0.96673. As the PDCSAP light curves are corrected for the dilution based on the positions of the stars over the observations, the behavior of the focal plane electronics, and the PRFs, we thus fixed the dilution to 1 (no dilution) for the modeling. 
The apertures of the ground-based photometery were not contaminated by neighboring stars, so we fixed the dilution factor to 1. 

To account for possible residual systematics and activity affecting the transit fit of both TESS Sectors 13 and 27, we tested fitting a Gaussian process (GP) using a Mat\'ern-3/2 kernel via {\sc \tt celerite}\xspace \citep{Foreman-Mackey2017} within the \juliet framework. A GP can account for correlated noise of various origins and propagates the uncertainty. It is particularly useful when we lack detailed knowledge on the origin of the noise and do not see any correlations that can be modeled in an uncomplicated manner. Therefore, a GP can correctly incorporate uncertainties introduced by red noise \citep[e.g.,][]{Gibson2012,Gibson2014}. Additionally, properly fitting out of transit data will more accurately set the baseline for in transit data. For both TESS sectors the GP fit displayed clear modulations, as shown in Figures \ref{fig:TESS13_full} and \ref{fig:TESS27_full}, and thus we included it in our final analysis. 

We also tested for correlated noise in the SAAO and EulerCam photometry by fitting the data with various combinations of polynomials in airmass, detector position, FWHM, and sky background by using a minimization of the Bayesian Information Criterion. We determined that no combination of linear or polynomial detrending for SAAO and EulerCam data significantly improved the light curve fit to warrant adding additional variables to the model, especially given the short out-of-transit baselines for both data sets. We exemplify this approach by displaying the polynomial detrending with sky background and airmass (the two parameters we found the most correlated with the SAAO and EulerCam) in Figure \ref{fig:groundphot_detrend}. With this detrending, we found a similar planetary radius value of $R_{p}$ = 1.00$\pm$0.04\,\rjup, which is within the uncertainites of the planetary radius derived without detrending. We also display the posterior distributions of these detrending parameters along with $p=R_p/R_{\star}$, showing that the planet radius is not correlated these parameters. We therefore did not perform any detrending for SAAO and EulerCam in our final analysis. We additionally tested the effect of the ground-based photometry on the planet radius by using only TESS photometry in a global model. With TESS only photometry we obtained a radius of $R_{p}$ = 1.02$\pm$0.04\,\rjup, again within the uncertainties of the planet radius used in the final global model that includes the ground-based photometry. We conclude that the ground-based photometry does not bias the results.

Figures \ref{fig:transits} and \ref{fig:rvs} display the final model fits to the photometry and RV data. In addition to all of the modeled parameters, Table \ref{tab:planet} also displays derived planet parameters including the inclination, impact parameter, semi-major axis, mass, and radius. We calculated the insolation using the equation:
\begin{equation}
    S [S_{\oplus}] = L_{*} [L_{\odot}] \left(a [\rm{AU}] \right)^{-2}.
\end{equation}
We calculated the equilibrium temperature assuming a Bond albedo of $A$\,=\,0.343 (the same as Jupiter's) and the semi-major axis distance $a$ using the equation:
\begin{equation}
T_{eq} = \teff (1-A)^{1/4}\sqrt{\frac{\rstar}{2a}}.
\end{equation}
We set upper and lower uncertainties for the equilibrium temperature by assuming Bond albedos of $A$\,=\,0 and $A$\,=\,0.686 (double that of  Jupiter), respectively.

\subsection{Activity and residual analysis} \label{sec:activity}

We also use the HARPS spectra to estimate the activity level of TOI-5542. The flux in the core of Ca II H \& K lines is well-known to estimate magnetic activity in solar-type stars, as demonstrated by the Mt Wilson program \citep[e.g.,][]{Wilson1978,Duncan1991,Baliunas1995}. The Mt Wilson `S index', which normalizes the measured Ca II core flux by the flux in two continuum bandpasses on the blue and red sides of the Ca II lines, has become a standard way of measuring chromospheric activity in stars. Here, we derive Mt. Wilson S indexes for TOI-5442 using HARPS spectra with the method detailed by \citet{Lovis2011}. However, to be able to compare the chromospheric flux alone (related to the energy that heats the chromosphere through the magnetic field) to other stars, the photospheric component has to be subtracted and the chromospheric flux normalized to the total (bolometric) luminosity of the star. \citet{Noyes1984} introduced the well-known quantity R'$_{HK}$, which can be determined from the B and V magnitudes and the Mt. Wilson S index \citep[e.g.,][]{Middelkoop1982,Noyes1984}. With our HARPS spectra we calculate log(R'$_{\rm{HK}}$) = $-$5.284 $\pm$ 0.263, which suggests that TOI-5542 is not a chromospherically active star, since stars with log(R'$_{\rm{HK}}$) $<$ $-$5.1 are generally considered very inactive \citep[e.g.,][]{Henry1996}.

To analyze the residuals, we ran a Bayesian analysis on the CORALIE and HARPS RV residuals with the software package \texttt{kima} \citep{Faria2018}. The RV time series is modeled with a sum of Keplerian signals. Different instruments are taken into account with individual offsets and jitters. We allowed our model to search for up to one Keplerian signal. To do so, we set the number of planets, which is a free parameter for \texttt{kima}, with a uniform prior between zero and one. The results of the search for one signal do not show any clear detection. As presented in \citet{Standing2022}, the samples from the posterior can be used to derive detection limits. The authors also show that this method is equivalent to the traditional test of injection and recovery of planetary signals. Figure \ref{fig:rv_resid} presents the posterior samples in the RV semi-amplitude versus orbital period space. The solid and dashed blue lines correspond to the RV signals of Jupiter, Saturn, and Neptune mass planets. We show that we are sensitive to sub-Saturn mass planets up to $\sim$100 days and for longer orbital periods (up to 300 days) we can exclude the presence of additional Jupiter mass planets.

\section{Discussion} \label{sec:discussion}

\begin{figure}
  \centering
  \includegraphics[width=0.49\textwidth]{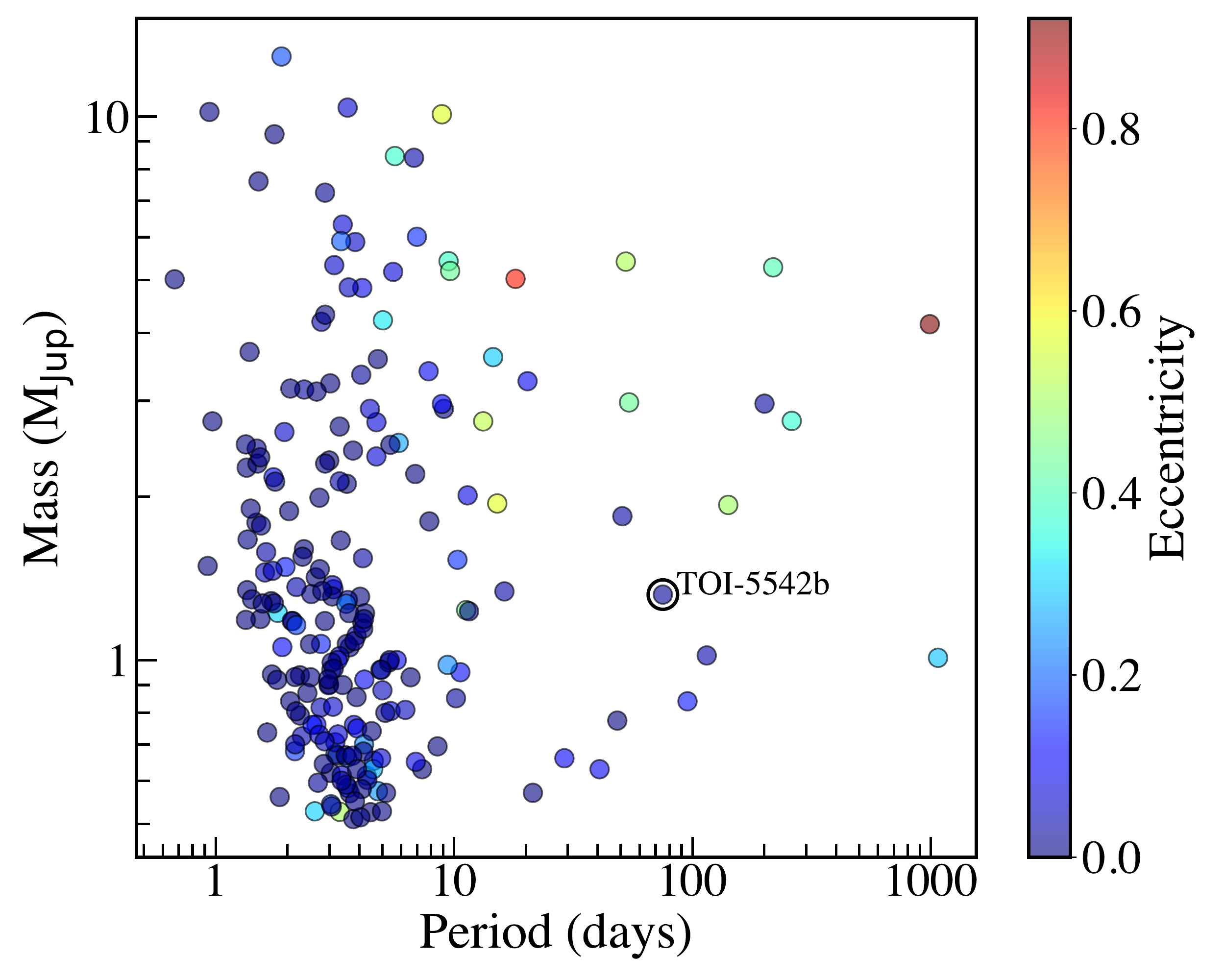}
 \caption{Mass as a function of orbital period for known giant planets (0.5\,\mjup $<$ $M_p$ $<$ 13\,\mjup) with well characterized masses ($\sigma_{M_p}$/$M_p$ $<$ 25\%) and radii ($\sigma_{R_p}$/$R_p$ $<$ 8\%) that have finite period and \teff values and stellar metallicity errors $<$0.25, consisting of 216 giant planets. The symbols of each planet are colored based on their eccentricity. TOI-5542b is marked with a black circle.}
  \label{fig:mass_per}
\end{figure}

We detected and characterized the warm Jupiter TOI-5542b that has a mass of $M_p$\,=\,1.32$^{+0.10}_{-0.10}$\,\mjup, radius of $R_p$\,=\,1.009$^{+0.036}_{-0.035}$\,\rjup, a period of 75.12 days, and a likely circular orbit with an eccentricity of 0.018$^{+0.026}_{-0.013}$. TOI-5542b is warm with an insolation of $S$\,=\,9.6$^{+1.0}_{-0.9}$\,$S_{\oplus}$ and equilibrium temperature of $T_{eq}$\,=\,441$^{+48}_{-74}$\,K. To put TOI-5542b into context we obtained planet parameters for known giant planets (0.5\,\mjup $<$ $M_p$ $<$ 13\,\mjup) from the NASA exoplanet archive\footnote{\url{https://exoplanets.nasa.gov/}} on 1 September 2022 and use 25\% mass and 8\% radius precision cutoffs in order to only consider well characterized planets with robust density measurements, as has been done in previous population studies \citep[e.g.,][]{Otegi2020}. We also required stellar metallicity values to have errors less than 0.25 and the planet to have a finite period and stellar \teff values in the archive. Our final sample of well-characterized giant planets consists of 216 planets, including TOI-5542b. We found 28 of these planets do not have defined eccentricities, but 25 of them have orbital periods less than 5 days, so we set the eccentricity to 0 and for the remaining three planets, we obtained estimates of their eccentricities from previous studies.

\subsection{Giant planet eccentricities}

We first compare TOI-5542b to our sample of well-defined giant planets by plotting mass as a function of the orbital period in Figure \ref{fig:mass_per}, which shows that TOI-5542b is one of the few well-characterized giant planets in the sparsely populated long-period regime. From Figure \ref{fig:mass_per}, we can see that giant planets exhibit a range of eccentricities and it appears that more massive ($M_{p}$ $\gtrsim$ 3-4 \mjup) planets appear to have a wider range of eccentricities than less massive giant planets, which was previously noted by \citet{Ribas2007}. Both hot and warm Jupiters are thought to have formed beyond the ice-line and then migrated inwards \citep[e.g.,][]{Bodenheimer2000,Rafikov2006}. Differences in the eccentricities of giant planets offer insights into different formation and migration pathways. 

Giant planets with lower eccentricities may have formed from disk-driven migration \citep[e.g.,][]{Goldreich1980,Ward1997,Baruteau2014} that tend to damp eccentricities \citep{Bitsch2013,Dunhill2013}. While giant planets with higher eccentricities may be the result of high-eccentricity migration from dynamical instabilities due to planet–planet gravitational interactions \citep[e.g.,][]{Rasio1996,Weidenschilling1996,LinIda1997} as well as secular perturbations due to distant stellar \citep[e.g.,][]{Wu2003,Fabrycky2007} or planetary companions \citep[e.g.,][]{Naoz2011,Wu2011}. However, none of these mechanisms can easily account for the observed warm Jupiter population and it is very likely a combination of these mechanisms. \citet{Petrovich2016} found $\sim$20\% of all warm Jupiters and most warm Jupiters with eccentricity $\gtrsim$0.4 are produced by high-eccentricity migration. The remaining population of low-eccentricity warm Jupiters ($e$ $\lesssim$ 0.2) likely obtained their current orbital configurations while the gaseous disk was still present \citep{Petrovich2016}, either by disk migration \citep[e.g.,][]{Goldreich1980} or in situ formation \citep[e.g.,][]{Batygin2016,Boley2016,Huang2016}.

We found a low eccentricity of 0.018$^{+0.026}_{-0.013}$ for TOI-5542b. We performed the test of \citet{Lucy1971}, where the statistical significance of the eccentric fit is given by:
\begin{equation}
    \rm{P}(e > 0) = 1 - \rm{exp}\left[ - \frac{\hat{e}}{2\sigma^{2}_{e}}\right],
\end{equation}
where $\hat{e}$ is the modal value of the eccentricity, which is well approximated by the median for a unimode distribution. We find the statistical significance of the eccentric fit to be P$(e > 0)$ = 0.68, which is well below the 5\% significance level suggested by \citet{Lucy1971}, and thus TOI-5542b likely has a circular orbit. With a circular orbit it is difficult to predict a formation or migration pathway for TOI-5542b, but given the absence of close planet companions based on the residuals of the RV and photometry data, it more likely formed via disk migration or in situ formation, compared to other mechanisms that are more likely to leave a planet with an eccentric orbit.

\begin{figure}
  \centering
  \includegraphics[width=0.49\textwidth]{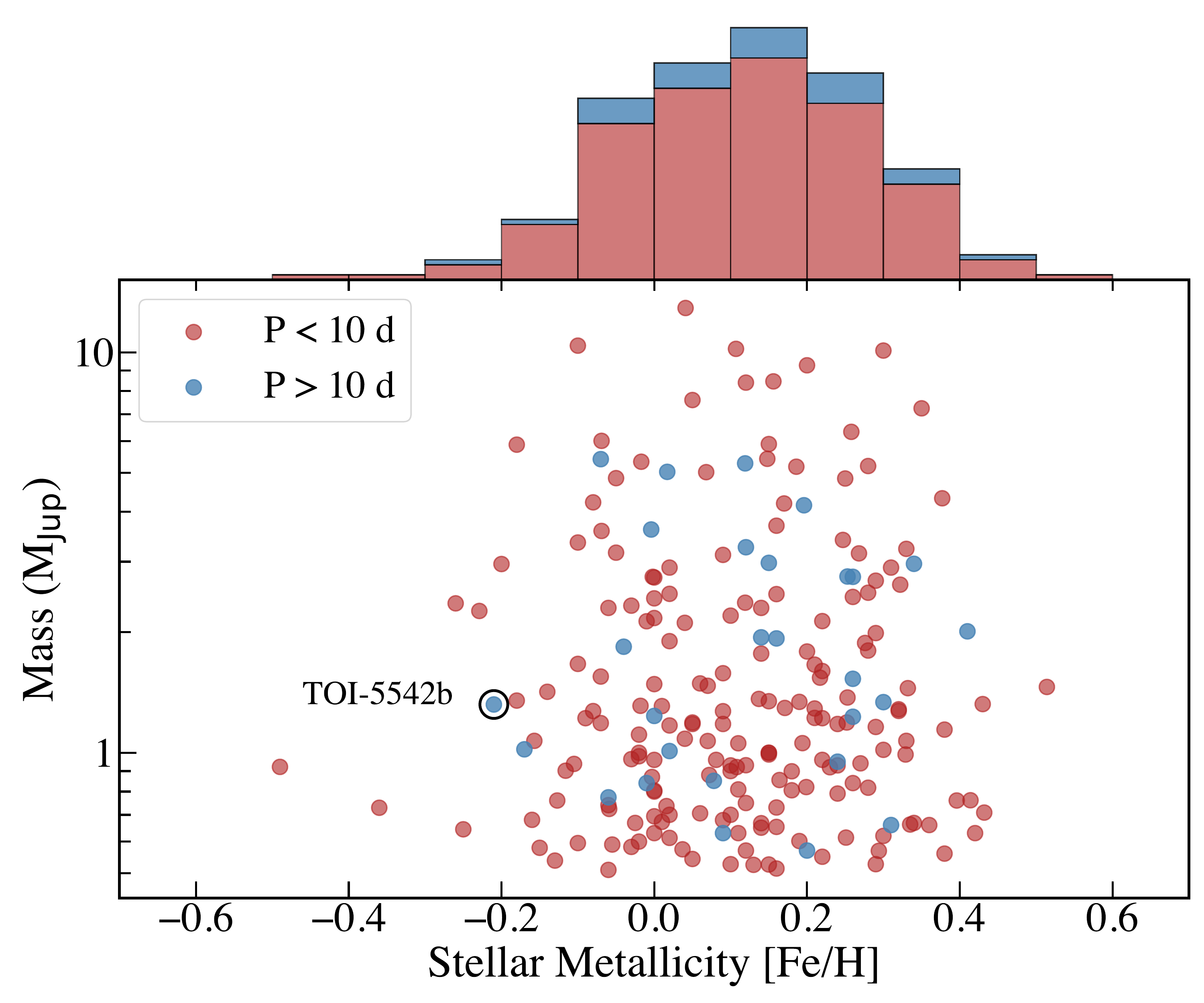}
 \caption{Mass as a function of host star metallicity for known giant planets (0.5\,\mjup $<$ $M_p$ $<$ 13\,\mjup) with well characterized masses ($\sigma_{M_p}$/$M_p$ $<$ 25\%) and radii ($\sigma_{R_p}$/$R_p$ $<$ 8\%) that have finite period and \teff values and stellar metallicity errors $<$0.25, consisting of 216 giant planets. The red symbols are planets with orbital periods $<$10 days (188 planets) and blue symbols have periods $>$10 days (28 planets). The histograms at the top show the metallicity distribution for the two populations. TOI-5542b is marked with a black circle.}
  \label{fig:mass_metal}
\end{figure}

\subsection{Host star metallicities of giant planet}

Since the early stages of exoplanet discoveries, studies have found that the occurrence rate of close-in giant planets is enhanced around stars with higher metallicity \citep[e.g.,][]{Gonzalez1997,Santos2004,Fischer2005,Johnson2010}. However, several recent studies have suggested that host star metallicity may also indicate the separation of two distinct giant planet populations at $\sim$3-4 \mjup, where more massive giant planets appear to have host stars with lower metallicities \citep[e.g.,][]{Santos2017,Schlaufman2018,Maldonado2019,GodaMatsuo2019}. These differences could serve as evidence of two different formation mechanisms, whereby lower mass giant planets more often form from core accretion \citep{Pollack1996} and higher mass giant planets may more often form via disk instability \citep{Boss1997} because disk instability is likely not to be as metallicity-dependent as core accretion \citep{Boss2002,Cai2006}. However, \citet{Adibekyan2019} did not find a transition point between two separate formation channels and thus suggested that high mass planets can form through different mechanisms depending on their initial environment. 

Notably, with a low metallicity of \feh\,=\,$-$0.21\,$\pm$\,0.08, TOI-5542  does not follow the traditional trend of high host-star metallicity for giant planets and does not bolster studies suggesting a difference among low- and high-mass giant planets. We place TOI-5542b into context in Figure \ref{fig:mass_metal} and plot planet mass as a function of host star metallicity for our sample of 216 well-defined giant planets. Although some recent studies have suggested that more massive giant planets appear not to have an inclination toward metal-rich stars, as compared to less massive giants, we note that in Figure \ref{fig:mass_metal} that there are no massive ($M_{p}$ $>$ 4\,\mjup) planets with metallicities lower than $-$0.2, in addition to there being almost no planets more massive than $\sim$\,1\,\mjup. \citet{Thorngren2016} also noted that more massive planets ($M_{p}$ $\gtrsim$ 2-3\,\mjup) are found far less often around low-metallicity stars when looking at a sample of 47 transiting warm giant planets with mass, radius, and age measurements. Additionally, when we split our sample of 216 well-defined giant planets at 4 \mjup we find that both samples are metal-rich with 27 massive giant planets having a mean host star \feh = 0.11\,$\pm$\,0.03 and 189 less massive giants having a mean of host star \feh = 0.11\,$\pm$\,0.01. We also split hot and warm giant planets at orbital periods of 10 days to see any differences in metallicites  -- but,  again, we found them both preferentially around metal-rich stars with a mean metallicity of \feh = 0.11\,$\pm$\,0.01 for the 188 hot Jupiters and \feh = 0.12\,$\pm$\,0.03 for the 28 warm (P $>$ 10 days) Jupiters in our sample.

\subsection{An old, warm Jupiter}

\begin{figure}
  \centering
  \includegraphics[width=0.49\textwidth]{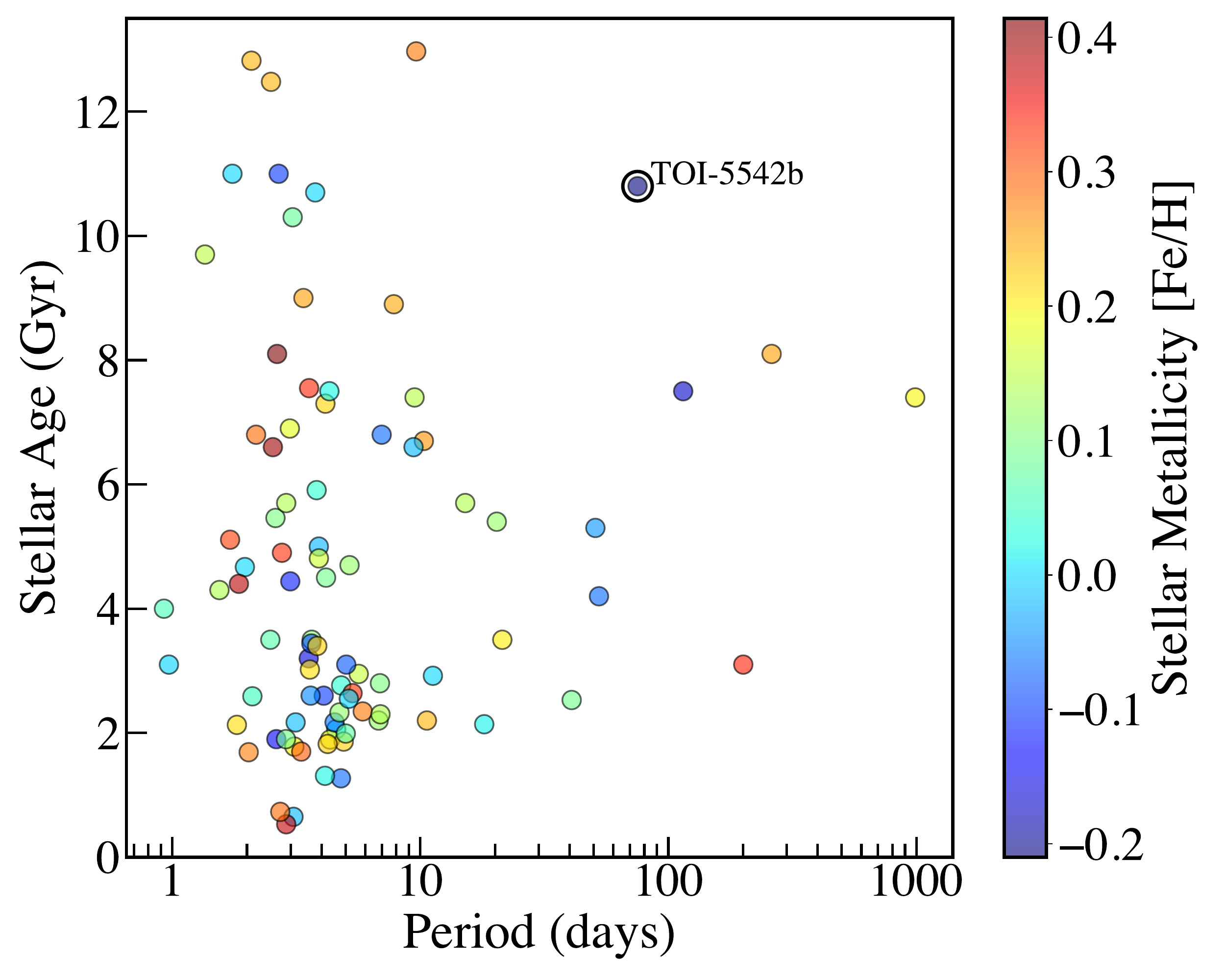}
 \caption{Stellar age as a function of orbital period for known giant planets (0.5\,\mjup $<$ $M_p$ $<$ 13\,\mjup) with well characterized masses ($\sigma_{M_p}$/$M_p$ $<$ 25\%) and radii ($\sigma_{R_p}$/$R_p$ $<$ 8\%) that have finite period and \teff values and stellar metallicity errors $<$0.25, as well as relatively well defined ages ($\sigma_{age}$/age $<$ 40\%) consisting of 87 giant planets. The symbols of each planet are colored based on their host star's metallicity. TOI-5542b is marked with a black circle.}
 \label{fig:age_period_metal}
\end{figure}

\begin{figure}
  \centering
  \includegraphics[width=0.49\textwidth]{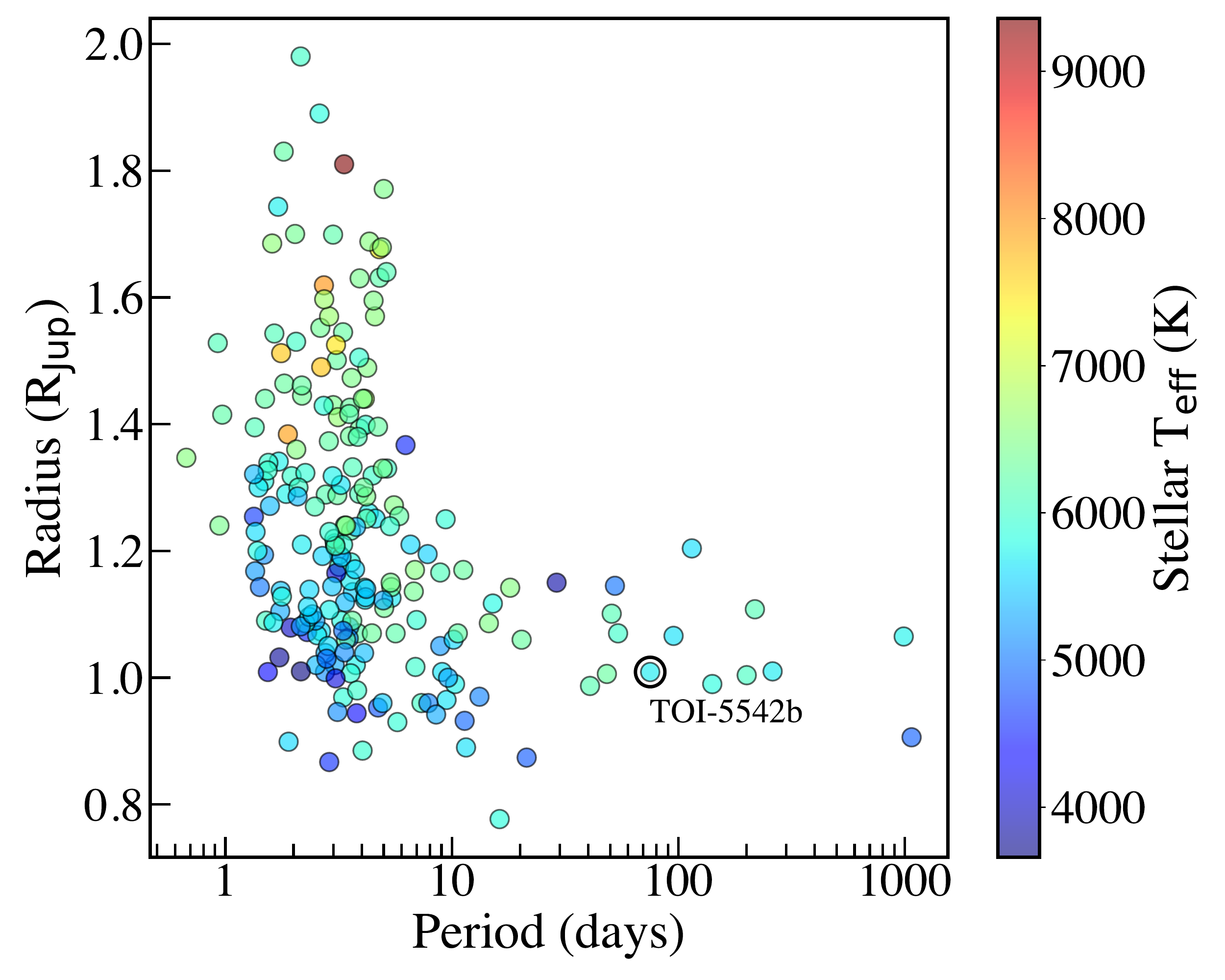}
 \caption{Radius as a function of orbital period for known giant planets (0.5\,\mjup $<$ $M_p$ $<$ 13\,\mjup) with well characterized masses ($\sigma_{M_p}$/$M_p$ $<$ 25\%) and radii ($\sigma_{R_p}$/$R_p$ $<$ 8\%) that have finite period and \teff values, and stellar metallicity errors $<$0.25, consisting of 216 giant planets. The symbols of each planet are colored based on their host star's \teff. TOI-5542b is marked with a black circle.}
  \label{fig:rad_period_teff}
\end{figure}

From our MIST evolutionary models (see Section \ref{sec:star_mist}), we find an age of $10.8^{+2.1}_{-3.6}$\,Gyr for TOI-5542, which is constrained from its low metallicity (\feh\,=\,$-$0.21\,$\pm$\,0.08), making it an old G dwarf. However, as discussed above, given its low eccentricity, TOI-5542b likely migrated early in the star system's history while the gaseous disk was still present. In Figure \ref{fig:age_period_metal}, we plot the host star age as a function of orbital period for giant planets in our sample that have relatively well defined ages ($\sigma_{age}$/age $<$ 40\%) consisting of 87 planets. Figure \ref{fig:age_period_metal} shows that TOI-5542b is one of the oldest known long-period warm Jupiters and one of the few with an age estimate.

In Figure \ref{fig:rad_period_teff}, we plot planet radius as a function of orbital period for our well-defined giant planet sample. Figure \ref{fig:rad_period_teff} displays how longer period warm Jupiters are not inflated from host star incident flux, which is known for planets receiving an incident flux below $\sim$2$\times$10$^{8}$ erg\,s${-1}$\,cm$^{-2}$ or $T_{eq}$ $\lesssim$ 1000\,K \citep{Demory2011,Miller2011}. This offers the advantage of being able to ignore inflation effects when constraining planet compositions. With an equilibrium temperature of $T_{eq}$ = 441$^{+48}_{-74}$\,K, TOI-5542b is well below this limit, making it a valuable data point for planet composition models and studies. 

As mentioned previously, we predicted the Rossiter-McLaughlin effect to be $\sim$26\,\ms\,for TOI-5542, which is large enough that the spin-orbit angle of the system can be measured with current high-resolution spectrographs. This obliquity measurement can help decipher the target’s dynamical past. For instance, in the case where a misaligned orbit were detected, the high-eccentricity migration scenario may appear to be a more likely scenario for its formation pathway \citep[e.g.,][]{Bourrier2018,Attia2021} but this may be unlikely given its circular orbit and the current lack of evidence for other planet companions.

\section{Conclusions} \label{sec:conclusion}

We report the discovery and characterization of the warm Jupiter TOI-5542b. The planet was first detected by TESS as two single transit events 375.6 days apart. We obtained radial velocities from the CORALIE and HARPS spectrographs, which constrained the orbital period of 75.12 days from the possible aliases of the two transits. We obtained a third transit simultaneously with the ground-based facilities SAAO, NGTS, and EulerCam, but we do not use NGTS in our global analysis. 

Using spectral analysis, SED fitting, and evolutionary models we measure the stellar parameters of TOI-5542 with \teff = $5700\pm80$\,K, \logg = $4.2\pm0.2$, \feh = $-0.21\pm0.08$, \mstar = $0.890^{+0.056}_{-0.031}$\,\msol, \rstar = $1.058\pm0.036$\,\rsol, and an age of $10.8^{+2.1}_{-3.6}$\,Gyr. We found TOI-5542b to have a mass of $M_p$ = 1.32$^{+0.10}_{-0.10}$\,\mjup, radius of $R_p$ = 1.009$^{+0.036}_{-0.035}$\,\rjup, and a likely circular orbit with an eccentricity of 0.018$^{+0.026}_{-0.013}$. TOI-5542b is warm with an insolation of $S$ = 9.6$^{+1.0}_{-0.9}$\,$S_{\oplus}$ and equilibrium temperature of $T_{eq}$ = 441$^{+48}_{-74}$\,K. TOI-5542b likely has a circular orbit and more likely formed via disk migration or in situ formation, rather than high-eccentricity migration mechanisms.  

We set TOI-5542b into context with a sample of well-characterized giant planets consisting of 216 planets including TOI-5542b. TOI-5542b is one of the few well-characterized giant planets in the sparsely populated long-period regime. With a low metallicity of \feh\,=\,$-$0.21\,$\pm$\,0.08, TOI-5542b does not follow the traditional trend of high host star metallicity for giant planets and does not bolster studies suggesting a difference in host star metallicities for low- and high-mass giant planets. When looking at our sample of well-characterized giant planets, we find both high-mass (4\,\mjup $<$ $M_{p}$ $<$ 13\,\mjup) and low-mass (0.5\,\mjup $<$ $M_{p}$ $<$ 4\,\mjup) giant planets both are preferentially located around metal-rich stars (mean \feh\,$>$\,0.1). We also found that both warm (P $>$ 10 days) and hot (P $<$ 10 days) Jupiters have mean metallicities above 0.1 dex. We determined that the Rossiter-McLaughlin effect for TOI-5542 would have an amplitude of $\sim$26\,\ms\,, which is high enough that the spin-orbit angle of the system can be measured with current high-resolution spectrographs. With an age of $10.8^{+2.1}_{-3.6}$\,Gyr, TOI-5542b is one of the oldest known warm Jupiters. With its equilibrium temperature of $T_{eq}$= 441$^{+48}_{-74}$\,K, TOI-5542b is not affected by inflation due to stellar incident flux, making it a valuable contribution in planetary composition and formation studies. 

\begin{acknowledgements}
We thank the Swiss National Science Foundation (SNSF) and the Geneva University for their continuous support to our planet low-mass companion search programs. This work was carried out in the frame of the Swiss National Centre for Competence in Research (NCCR) $PlanetS$ supported by the Swiss National Science Foundation (SNSF). This publication makes use of The Data \& Analysis Center for Exoplanets (DACE), which is a facility based at the University of Geneva (CH) dedicated to extrasolar planet data visualization, exchange, and analysis. DACE is a platform of NCCR $PlanetS$ and is available at https://dace.unige.ch. This paper includes data collected by the TESS mission. Funding for the TESS mission is provided by the NASA Explorer Program. We acknowledge the use of public TESS data from pipelines at the TESS Science Office and at the TESS Science Processing Operations Center. Resources supporting this work were provided by the NASA High-End Computing (HEC) Program through the NASA Advanced Supercomputing (NAS) Division at Ames Research Center for the production of the SPOC data products. This paper uses observations made
at the South African Astronomical Observatory (SAAO). The {\sc Starlink} software \citep{Currie2014} is currently supported by the East Asian Observatory.
MNG acknowledges support from the European Space Agency (ESA) as an ESA Research Fellow.
AJ acknowledges support from ANID -- Millennium  Science  Initiative -- ICN12\_009 and from FONDECYT project 1210718. M.L. acknowledges support of the Swiss National Science Foundation under grant number PCEFP2\_194576.
NG thanks Tara Grieves for her support, passion, and kindness over the years.

\end{acknowledgements}

%
%

\bibliographystyle{aa}
\bibliography{bib}

\begin{appendix} 
\section{Supplementary material}

\begin{figure*}
  \centering
  \includegraphics[width=0.95\textwidth]{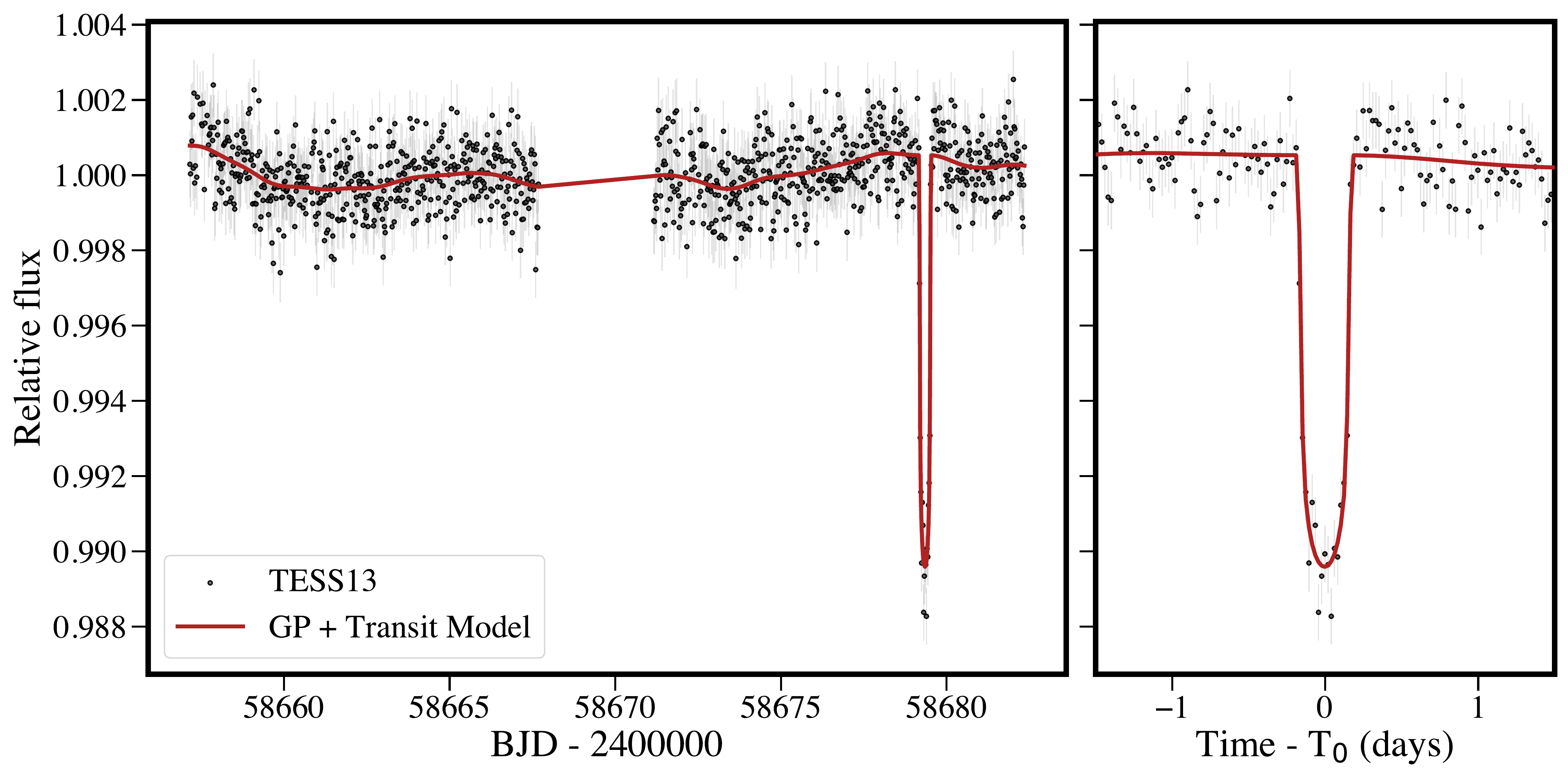}
 \caption{Full TESS Sector 13 data of TOI-5542 . Zoom-in on transit shown on the right. The black circles display the 30-minute SPOC data used in the analysis and the red line displays the full GP and transit model.}
  \label{fig:TESS13_full}
\end{figure*}

\begin{figure*}
  \centering
  \includegraphics[width=0.95\textwidth]{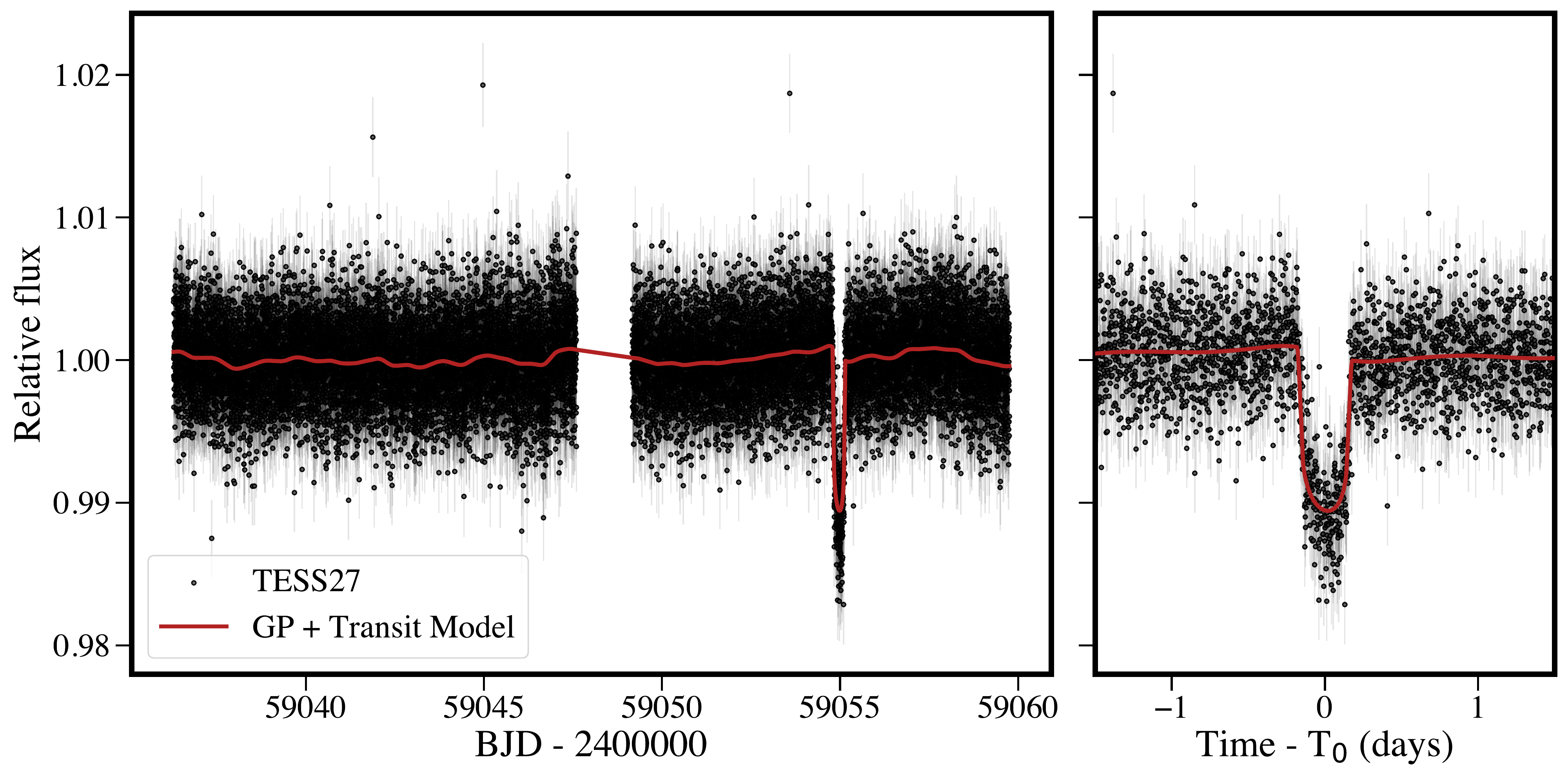}
 \caption{\textit{Left}: Full TESS Sector 27 data of TOI-5542 . Zoom-in on transit shown
on the right. The black circles display the 2-minute SPOC data used in the analysis and the red line displays the full GP and transit model.}
  \label{fig:TESS27_full}
\end{figure*}

\begin{figure*}
  \centering
  \includegraphics[width=0.95\textwidth]{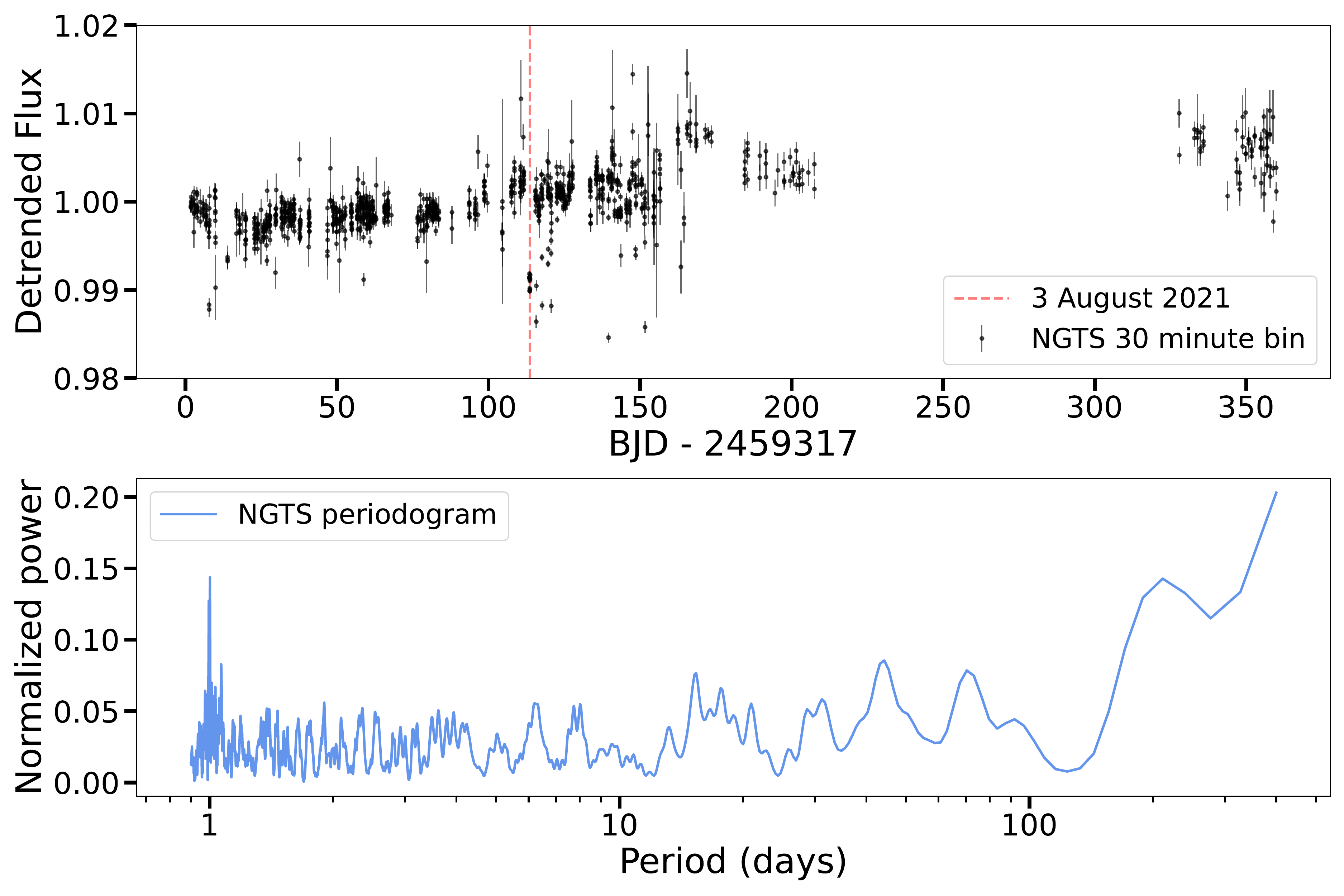}
 \caption{ NGTS data binned to 30 minutes (top). The dashed red line displays the night TOI-5542b transited and shows the relative offset compared to nearby nights. Due to overall variation of the light curve and no ingress or egress of the transit, we do not include NGTS data in our global model. Lomb-Scargle periodogram of NGTS data (bottom). We do not see any significant peaks corresponding to a likely stellar roation period.}
  \label{fig:ngts_full}
\end{figure*}

\begin{figure*}
  \centering
  \includegraphics[width=0.48\textwidth]{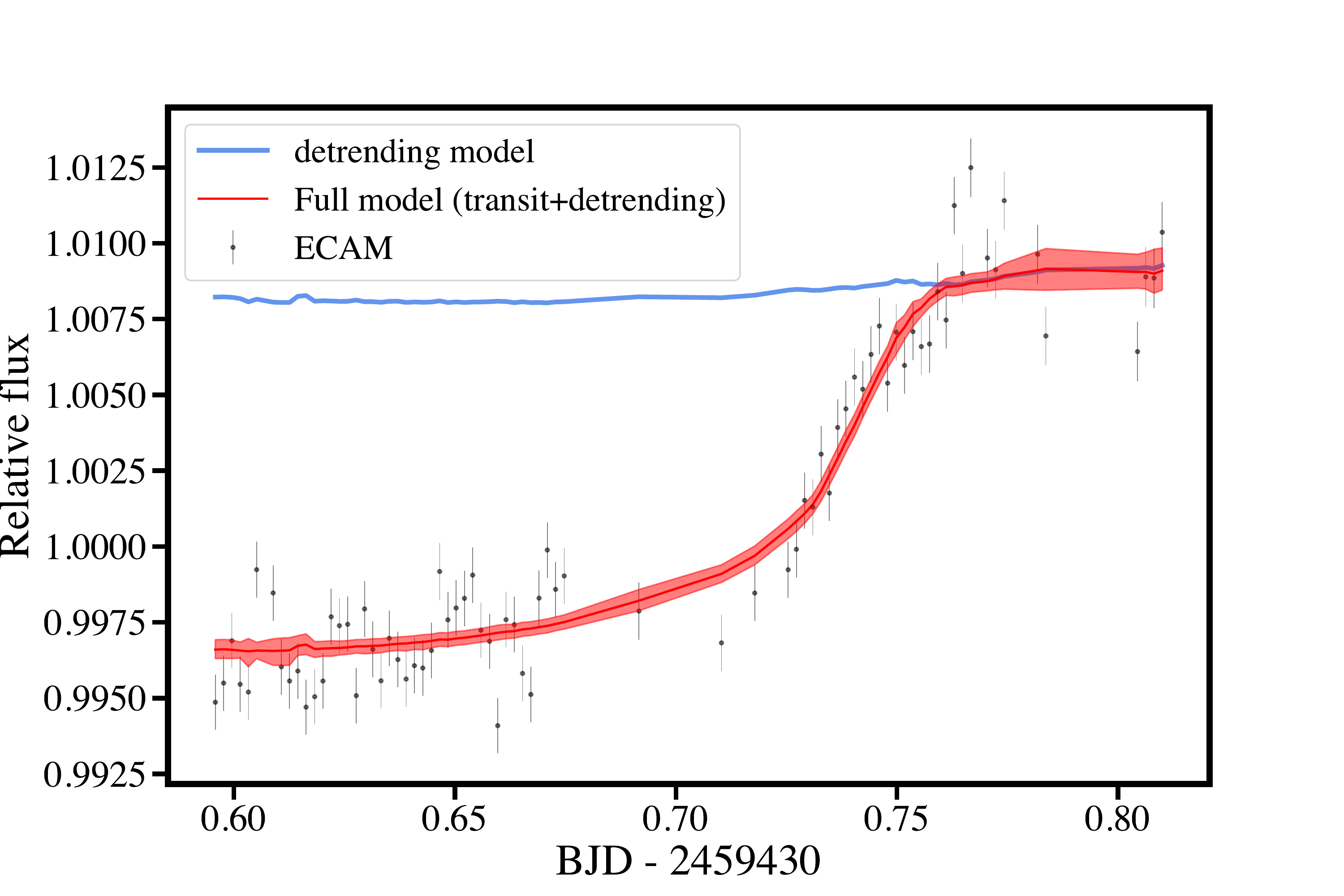}
  \includegraphics[width=0.48\textwidth]{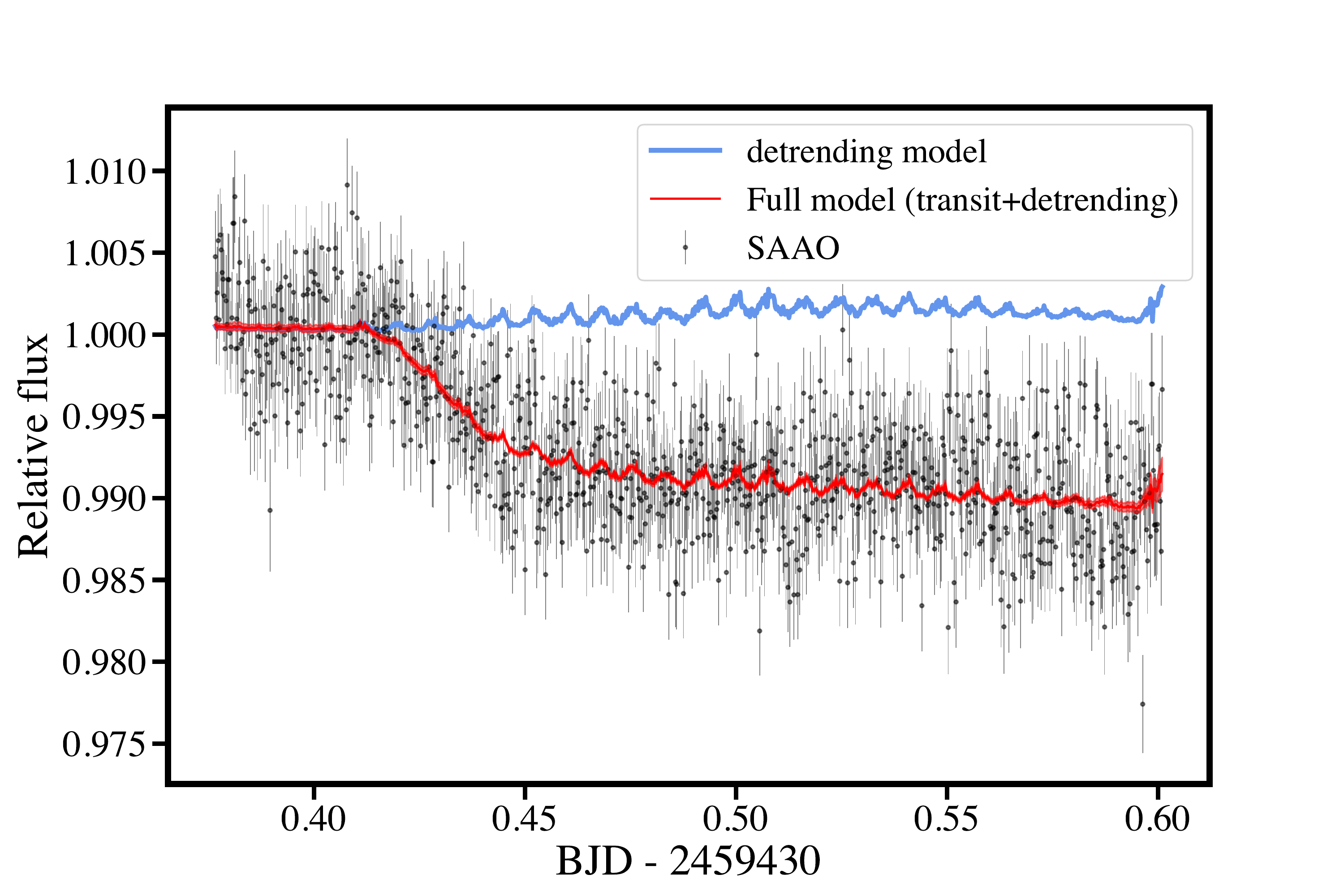}
  \includegraphics[width=0.95\textwidth]{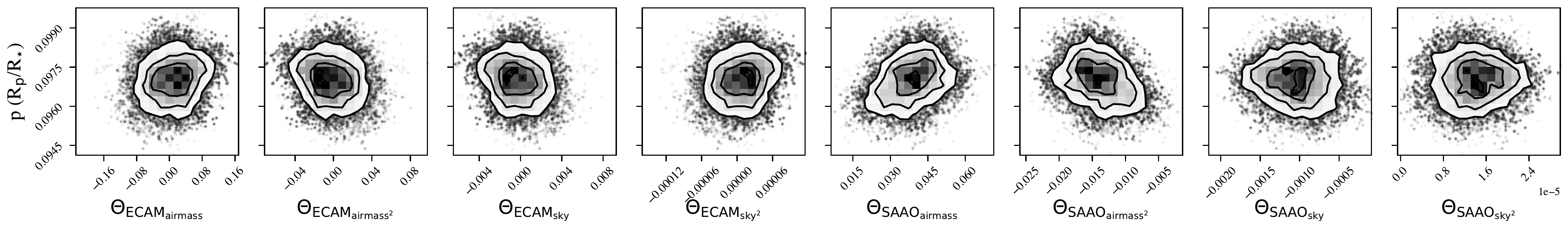}
 \caption{Detrending fits for the ground-based photometry. The top plots display the fits obtained for EulerCam and SAAO when detrending with polynomials in sky and airmass. The bottom plot displays the posterior distributions of these detrending parameters along with $p=R_p/R_{\star}$, which shows the planet radius is not correlated these parameters. We obtained a similar planet radius  of $R_{p}$ = 1.00$\pm$0.04\,\rjup with this fit that was within the uncertainites of the planet radius derived without detrending and this value used for the final model.}
  \label{fig:groundphot_detrend}
\end{figure*}

\begin{figure*}
  \centering
  \includegraphics[width=0.7\textwidth]{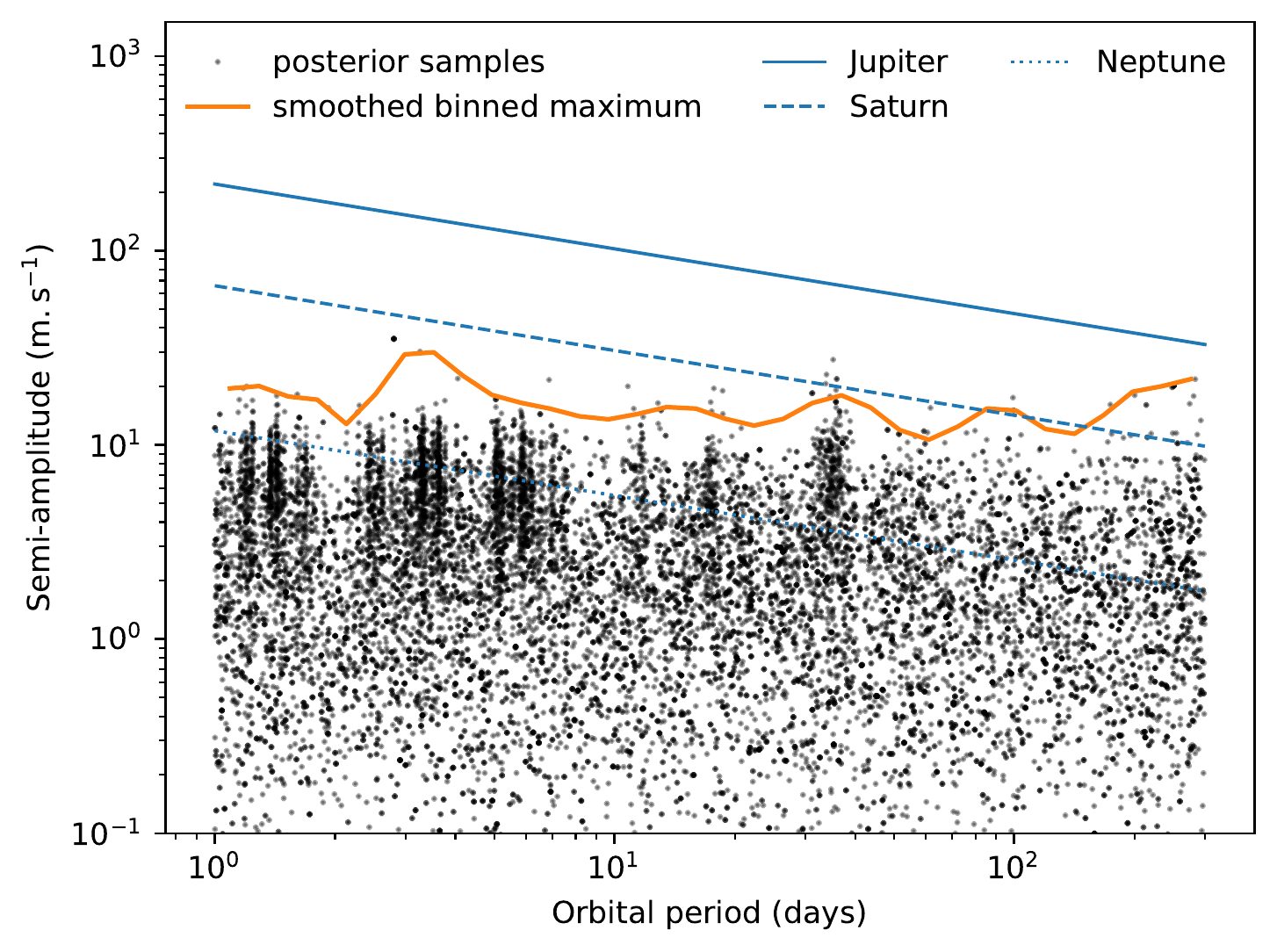}
 \caption{RV residuals analysis presenting the posterior samples in the RV semi-amplitude versus orbital period space. The results of the search for a second companion do not show any clear detection. The solid and dashed blue lines correspond to the RV signals of Jupiter-, Saturn-, and Neptune-mass planets. We show that we are sensitive to sub-Saturn-mass planets up to $\sim$100 days and for longer orbital periods (up to 300 days), we can exclude the presence of additional Jupiter-mass planets.}
  \label{fig:rv_resid}
\end{figure*}

\begin{table}
\footnotesize
    \centering
    \begin{tabular}{ccc}
        & CORALIE & \\ 
        \hline\hline
        Time [BJD TDB] & RV [$\ms$] & RV error [$\ms$] \\
        \hline
2459306.87282241 & -54451.2  & 83.5 \\ 
2459316.88105868 & -54461.9  & 63.2 \\ 
2459322.86463654 & -54353.4  & 57.9 \\ 
2459329.90496341 & -54357.1  & 60.9 \\ 
2459340.88479492 & -54359.8  & 175.5 \\ 
2459350.88006593 & -54445.1  & 123.6 \\ 
2459358.78809994 & -54462.4  & 51.9 \\ 
2459365.76126559 & -54486.7  & 39.6 \\ 
2459372.72588852 & -54499.1  & 48.3 \\ 
2459379.72408383 & -54444.7  & 51.8 \\ 
2459408.59341530 & -54291.4  & 101.3 \\ 
2459415.62714956 & -54393.7  & 69.6 \\ 
2459429.68413281 & -54362.4  & 46.0 \\ 
2459434.64777952 & -54318.1  & 103.0 \\ 
2459438.73530854 & -54458.8  & 74.7 \\ 
2459447.75931134 & -54483.5  & 35.6 \\ 
2459454.60669680 & -54445.3  & 38.4 \\ 
2459459.53970780 & -54457.5  & 90.1 \\ 
2459464.51671181 & -54429.8  & 64.2 \\ 
2459468.67790552 & -54458.3  & 88.9 \\ 
2459473.60548228 & -54410.3  & 42.2 \\ 
2459476.57988364 & -54355.9  & 52.2 \\ 
2459480.53595794 & -54293.8  & 51.5 \\ 
2459490.60161776 & -54333.8  & 44.3 \\ 
2459497.53153252 & -54267.3  & 81.9 \\ 
2459504.50552615 & -54390.5  & 58.2 \\ 
2459509.57621809 & -54473.8  & 57.1 \\ 
2459513.57881157 & -54374.4  & 52.7 \\ 
2459525.53964922 & -54523.6  & 76.6 \\ 
2459532.54785833 & -54474.2  & 54.4 \\ 
2459541.52832420 & -54393.1  & 67.6 \\ 
\\
        & HARPS & \\ 
        \hline\hline
        Time [BJD TDB] & RV [$\ms$] & RV error [$\ms$] \\
        \hline
2459422.67412022 & -54343.21  & 9.51 \\ 
2459423.68107993 & -54322.18  & 7.07 \\ 
2459424.65831681 & -54326.17  & 5.35 \\ 
2459425.61910614 & -54359.24  & 8.40 \\ 
2459428.76938868 & -54365.23  & 5.69 \\ 
2459430.64989303 & -54390.56  & 10.10 \\ 
2459432.61062220 & -54401.89  & 7.22 \\ 
2459433.68498673 & -54399.99  & 12.35 \\ 
2459449.67700834 & -54431.81  & 25.54 \\ 
2459460.60745780 & -54410.36  & 6.44 \\ 
2459461.64402816 & -54412.78  & 7.86 \\ 
2459462.59750084 & -54409.14  & 6.85 \\ 
2459463.63573612 & -54400.62  & 5.59 \\ 
2459464.55511221 & -54404.66  & 11.84 \\ 
2459498.60139309 & -54340.57  & 10.14 \\ 
2459521.54616908 & -54440.69  & 6.02 \\ 
2459529.51177399 & -54429.92  & 6.50 \\ 
2459539.54222557 & -54402.57  & 7.71 \\ 

        \hline
        \hline
    \end{tabular}
    \caption{Radial velocities of TOI-5542 from the CORALIE and HARPS spectrographs.}
    \label{tab:rv}
\end{table}

\end{appendix}

\end{document}